\documentclass[twocolumn,pra,longbibliography,superscriptaddress]{revtex4-2}
\usepackage[latin1]{inputenc}
\usepackage{graphicx}
\usepackage{amsmath}
\usepackage{amsthm}
\usepackage{bm}
\usepackage{layout}
\usepackage{float}
\usepackage{amsfonts}
\usepackage{amssymb}%
\usepackage[margin=0.75in]{geometry}
\usepackage{color}
\usepackage{soul}
\usepackage{mathtools}
\usepackage[colorlinks=true,citecolor=blue]{hyperref}
\theoremstyle{definition}

\usepackage[capitalise,compress]{cleveref}

\newcommand{\ket}[1]{\left | #1 \right\rangle}

\newcommand{\bra}[1]{\left \langle #1 \right |}

\newcommand{\abs}[1]{\left | #1 \right|}
\renewcommand{\epsilon}{\varepsilon}
\renewcommand{\O}[1]{ O\left(#1\right)}

\newcommand{\norm}[1]{\left\|#1\right\|}

\newcounter{para}
\newcommand{\dist}{\textrm{dist}}

\makeatletter
\newcommand*\bigcdot{\mathpalette\bigcdot@{.5}}
\newcommand*\bigcdot@[2]{\mathbin{\vcenter{\hbox{\scalebox{#2}{$\m@th#1\bullet$}}}}}

\makeatother

\usepackage{array}
\newcolumntype{L}{>{$}l<{$}} % math-mode version of "l" column type
\newcolumntype{C}{>{$}c<{$}} % math-mode version of "c" column type
\newcolumntype{R}{>{$}r<{$}} % math-mode version of "r" column type

\usepackage{xcolor}

\usepackage{mathtools}

\usepackage{xr}
\makeatletter
\newcommand*{\addFileDependency}[1]{% argument=file name and extension
  \typeout{(#1)}
  \@addtofilelist{#1}
  \IfFileExists{#1}{}{\typeout{No file #1.}}
}
\makeatother

\usepackage{tikz}

\usepackage{mdframed}
\newmdenv[topline=false,rightline=false,bottomline=false,linewidth=2pt,linecolor=white!60!black,]{leftborder}

\usepackage{qcircuit}

\usepackage{booktabs}
\newcolumntype{C}{>{$}c<{$}}
\AtBeginDocument{
\heavyrulewidth=.08em
\lightrulewidth=.05em
\cmidrulewidth=.03em
\belowrulesep=.65ex
\belowbottomsep=0pt
\aboverulesep=.4ex
\abovetopsep=0pt
\cmidrulesep=\doublerulesep
\cmidrulekern=.5em
\defaultaddspace=0.5em
}
\newcommand{\GHZ}{\ket{\text{GHZ}}}
\newcommand{\GHZab}{\ket{\text{GHZ}(a,b)}}

\newcommand{\polylog}{\text{polylog}}

\usepackage[english]{babel}
\makeatletter
\let\ORIbbl@fixname\bbl@fixname
\def\bbl@fixname#1{%
  \@ifundefined{languagealias@\expandafter\string#1}
    {\ORIbbl@fixname#1}
    {\edef\languagename{\@nameuse{languagealias@#1}}}%
}
\newcommand{\definelanguagealias}[2]{%
  \@namedef{languagealias@#1}{#2}%
}
\makeatother

\definelanguagealias{en}{english}
\definelanguagealias{EN}{english}

\AtBeginDocument{%
    \newwrite\bibnotes
    \def\bibnotesext{Notes.bib}
    \immediate\openout\bibnotes=\jobname\bibnotesext
    \immediate\write\bibnotes{@CONTROL{REVTEX41Control}}
    \immediate\write\bibnotes{@CONTROL{%
    apsrev41Control,author="08",editor="1",pages="1",title="0",year="1"}}
     \if@filesw
     \immediate\write\@auxout{\string\citation{apsrev41Control}}%
    \fi
}%

\usepackage{amstext} % for \text macro
\usepackage{array}   % for \newcolumntype macro
\newcolumntype{L}{>{$}l<{$}} % math-mode version of "l" column type

\newcommand{\mysupcite}[1]{^{\tiny [\text{\citenum{#1}}]}}
\begin{document}
%%%%%%%%%%%%%%%%%%%%%%%%%%%%%%%%%%%%%%%%%%%%%
\title{\vspace*{-0.3in} Optimal State Transfer and Entanglement Generation in Power-law Interacting Systems}
\date{\today}
 
\author{Minh C. Tran}
\affiliation{Joint Center for Quantum Information and Computer Science,
NIST/University of Maryland, College Park, MD 20742, USA}
\affiliation{Joint Quantum Institute, NIST/University of Maryland, College Park, MD 20742, USA}

\author{Abhinav Deshpande}
\author{Andrew Y. Guo}
\affiliation{Joint Center for Quantum Information and Computer Science,
NIST/University of Maryland, College Park, MD 20742, USA}
\affiliation{Joint Quantum Institute, NIST/University of Maryland, College Park, MD 20742, USA}

\author{Andrew Lucas}
\affiliation{Department of Physics, University of Colorado, Boulder CO 80309, USA}
\affiliation{Center for Theory of Quantum Matter, University of Colorado, Boulder CO 80309, USA}

\author{Alexey V. Gorshkov}
\affiliation{Joint Center for Quantum Information and Computer Science,
NIST/University of Maryland, College Park, MD 20742, USA}
\affiliation{Joint Quantum Institute, NIST/University of Maryland, College Park, MD 20742, USA}

\begin{abstract}
	We present an optimal protocol for encoding an unknown qubit state into a
	multiqubit Greenberger-Horne-Zeilinger-like state and, consequently, transferring quantum information in large systems exhibiting power-law ($1/r^\alpha$) interactions.
	For all power-law exponents $\alpha$ between $d$ and $2d+1$, where $d$ is the dimension of the system,
	the protocol yields a polynomial speedup for $\alpha>2d$ and a superpolynomial speedup for $\alpha\leq 2d$, compared to the state of the art.
	For all $\alpha>d$, the protocol saturates the Lieb-Robinson bounds (up to subpolynomial corrections), thereby establishing the optimality of the protocol and the tightness of the bounds in this regime.
	The protocol has a wide range of applications, including in quantum sensing, quantum computing, and preparation of topologically ordered states.
	In addition, the protocol provides a lower bound on the gate count in digital simulations of power-law interacting systems.
\end{abstract}
\maketitle
%%%%%%%%%%%%%%%%%%%%%%%%%%%%%%%%%%%%%%%%%%%%%
\section{Introduction}
Harnessing entanglement between many particles is key to a quantum advantage in applications including sensing and time-keeping~\cite{winelandSpinSqueezingReduced1992,bollingerOptimalFrequencyMeasurements1996a}, secure communication~\cite{wehnerQuantumInternetVision2018}, and quantum computing~\cite{feynmanSimulatingPhysicsComputers1982,shorAlgorithmsQuantumComputation1994}.
For example, encoding quantum information into a multiqubit  Greenberger-Horne-Zeilinger-like (GHZ-like) state
is particularly desirable as a subroutine in many quantum applications,  including metrology \cite{bollingerOptimalFrequencyMeasurements1996a}, quantum computing \cite{pham13,guoImplementingFastUnbounded2020}, anonymous quantum communication \cite{christandl05,brassard09}, and quantum secret sharing \cite{hilleryQuantumSecretSharing1999}.

The speed at which one can unitarily encode an unknown qubit state $a \ket{0} + b \ket{1}$ into a GHZ-like state $a \ket{00 \dots 0} + b \ket{11 \dots 1}$ of a large system is constrained by Lieb-Robinson bounds~\cite{LR,NachtergaeleOS2006,Nachtergaele2006,HK,GongFF,Foss-FeigG,Storch15,NRSS09,SHKM10,SH10,tranLocalityDigitalQuantum2019a,elseImprovedLiebRobinsonBound2018,chenFiniteSpeedQuantum2019,tranHierarchyLinearLight2020a,kuwaharaStrictlyLinearLight2020} and depends on the nature of the interactions in the system.
In systems with finite-range interactions and power-law interactions decaying with distance $r$ as $1/r^\alpha$ for all $\alpha\geq 2d+1$, where $d$ is the dimension of the system, the Lieb-Robinson bounds imply a linear light cone for the propagation of quantum information~\cite{chenFiniteSpeedQuantum2019,kuwaharaStrictlyLinearLight2020}.
Consequently, in such systems, the linear size of a GHZ-like state that can be prepared from unentangled particles cannot grow faster than linearly with time.

\begin{figure} [ht]
\includegraphics[width=0.41\textwidth]{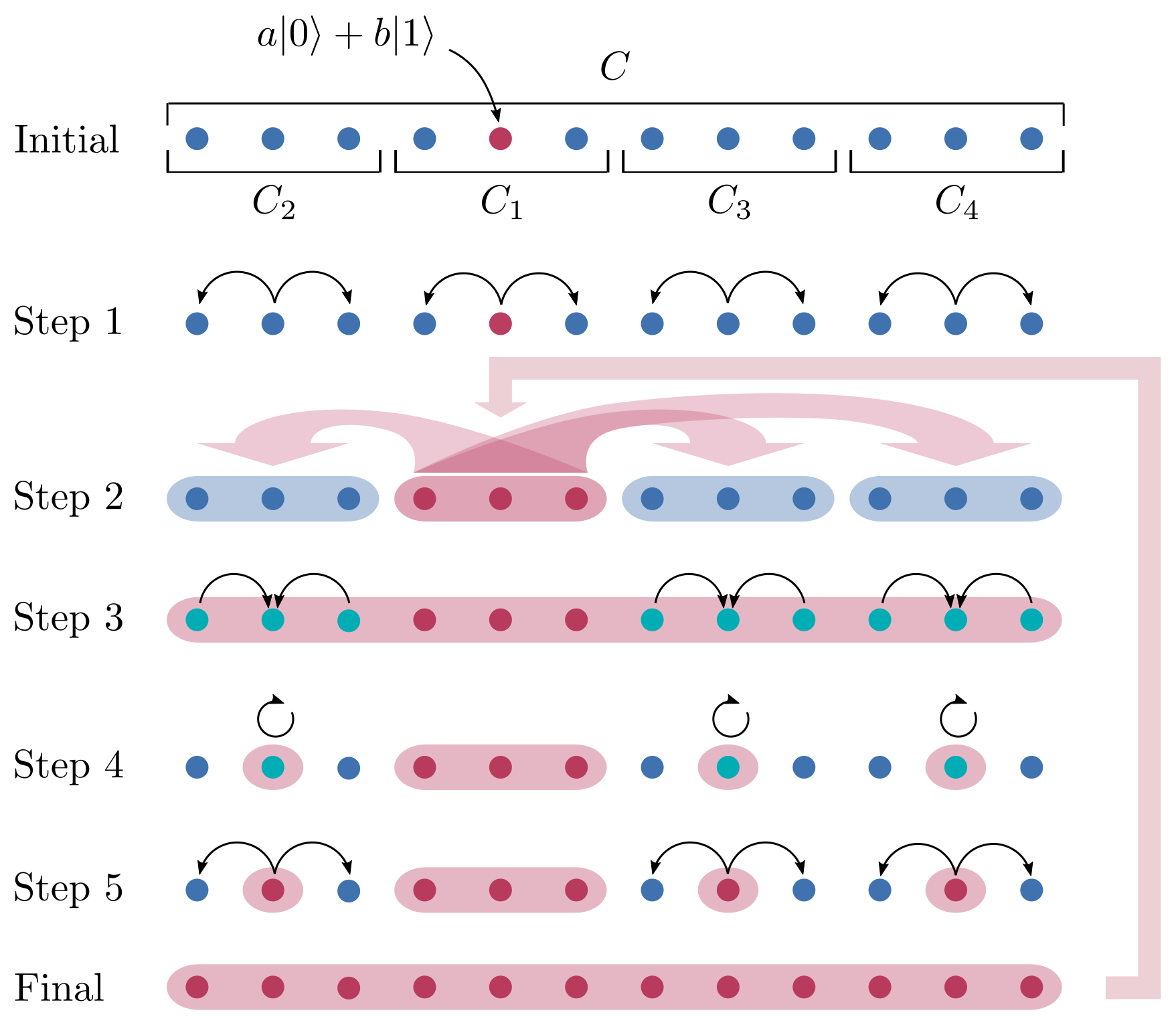}
\caption{A demonstration of our protocol for encoding a qubit into a GHZ-like state in a one-dimensional system $C$.
Initially, the unknown coefficients $a,b$ are encoded in one qubit (red circle) while the other qubits are each initialized in state $\ket{0}$.
The first step of the protocol assumes the ability to encode information into GHZ-like states in subsystems $C_1,\dots,C_4$ using, for example, nearest-neighbor interactions.
In step 2, we apply a generalized controlled-PHASE gate [\cref{eq:cPhase}] between the subsystems to ``merge'' the GHZ-like states into an entangled state between all sites. The last three steps rotate this entangled state into the desired GHZ-like state by concentrating the entanglement in each subsystem onto one qubit, applying single-qubit rotations, and redistributing the entanglement to the rest of the system.
Repeatedly feeding the resulting GHZ-like state back into step 2 of the protocol yields larger and larger GHZ-like states.
}
\label{fig:protocol}
\end{figure}

The Lieb-Robinson bounds become less stringent for longer-range interactions, i.e.\ those with $\alpha < 2 d+1$.
The bounds theoretically allow quantum information to travel a distance $r$ in time $t$ that scales sublinearly with~$r$~\cite{HK,GongFF,Foss-FeigG,tranLocalityDigitalQuantum2019a}.
However, no protocol in the present literature can saturate these bounds.
In particular, existing protocols for $\alpha \in (d,2d]$ are exponentially slower than what is allowed by the corresponding bounds.
Up until now, the existence of this gap between the Lieb-Robinson bounds and the achievable protocols has meant that at least one of the two is not yet optimal, hinting at either a tighter Lieb-Robinson bound or the possibility of speeding up many quantum information processing tasks.

In this paper, we close the gap for all $\alpha\in (1,3)$ in one dimension and $\alpha\in (d,2d]$ in $d>1$ dimensions by designing a protocol for encoding an arbitrary qubit into a multiqubit GHZ-like state and, subsequently, transferring information at the limits imposed by the Lieb-Robinson bounds.
There are three key implications of the protocol.
First, within these regimes of $\alpha$, it establishes the tightness of the Lieb-Robinson bounds, up to subpolynomial corrections, and effectively puts an end to the fifteen-year search for a tighter bound.
The scaling of time with the size of the GHZ states in our protocol for $\alpha\in (2d,2d+1)$ coincides with the conjectured generalization of the light cone in Ref.~\cite{chenFiniteSpeedQuantum2019} to $d>1$ dimensions and, therefore, provides strong evidence for the conjecture.
Second, our protocol implies optimal designs for future experiments on power-law interacting systems, including trapped ions~\cite{Kim2011,britton12}
($\alpha\in [0,3]$)
in one and two dimensions~\footnote{Quantum computing with trapped ions usually uses resonant addressing and real excitations of the motional modes~\cite{debnath_demonstration_2016}. On the other hand, for one-dimensional chains of trapped ions, the off-resonant addressing scheme, which results in spin models with tunable approximately-power-law couplings ($\alpha\in [0,3]$), is popular among recent analog quantum simulation experiments (For example, see Ref.~\cite{Pagano25396}).}, ultracold atoms in photonic crystals~\cite{Douglas2015,gonzalez-tudela15}, van-der-Waals interacting Rydberg atoms~\cite{Saffman10,bernien17} ($\alpha = 6$) in three dimensions \cite{barredo18}, as well as the very common case of dipolar interactions in nitrogen-vacancy centers~\cite{Maze2011}, polar molecules \cite{yan13}, and dipole-dipole interacting Rydberg atoms \cite{leseleuc18b} ($\alpha = 3$) in two dimensions.
Finally, our protocol implies a lower bound on the gate count in simulating power-law interacting systems on a quantum computer, providing a benchmark for the performance of quantum simulation algorithms.

The structure of the paper is as follows.
In \cref{sec:setup}, we define our setting and introduce the main result: the optimal state-transfer time in power-law interacting systems~[\cref{eq:Tbound}]. 
In \cref{sec:protocol}, we describe the corresponding optimal protocol for generating entanglement and subsequently transferring quantum information.
At the end of \cref{sec:protocol}, we discuss the key ingredients that make the protocol outperform previously known protocols. 
Readers who are interested in the conceptual implications of the protocol may also skip ahead to \cref{sec:discussion}, where we establish the tightness of existing Lieb-Robinson bounds and discuss implications for other types of speed limits associated with quantum information propagation.

\section{Setup and Results}\label{sec:setup}
We first describe the setting of the problem and the main result in this section.
For simplicity, we consider a $d$-dimensional hypercubic lattice $\Lambda$ and a two-level system located at every site of the lattice.
Our protocol generalizes straightforwardly to all regular lattices.
Without loss of generality, we assume that the lattice spacing is one.
We consider a power-law interacting Hamiltonian
	$H(t) = \sum_{i,j\in \Lambda} h_{ij}(t), $
where $h_{ij}(t)$ is a Hamiltonian supported on sites $i,j$ such that, at all times $t$ and for all $i \neq j$, we have $\norm{h_{ij}}\leq 1/\dist({i,j})^\alpha$, where $\dist({i,j})$ is the distance between $i,j$, $\norm{\cdot}$ is the operator norm, and $\alpha\geq 0$ is a constant.
We use $\GHZab_S$ to denote the GHZ-like state over sites in $S\subseteq \Lambda$:
\begin{align}
	\GHZab_S
	\equiv a\ket{\bar0}_S + b\ket{\bar 1}_S,
\end{align}
where $\ket{\bar x}_S \equiv \bigotimes_{j\in S} \ket{x}_j$ $(x = 0,1)$ are product states over all sites in $S$ and $a,b$ are complex numbers such that $\abs{a}^2+\abs{b}^2 = 1$.
In particular, we use $\GHZ$ to denote the symmetric state $a = b = 1/\sqrt2$.

Given a $d$-dimensional hypercube $C\subseteq \Lambda$ of length $r\geq 1$, we consider the task of encoding a possibly unknown state $a\ket{0}+b\ket{1}$ of a site $c \in C$ into the GHZ-like state $\GHZab_C$ over $C$, assuming that all sites in $C$, except for $c$, are initially in the state $\ket{0}$.
Specifically, we construct a time-dependent, power-law interacting Hamiltonian $H(t)$ that generates $U(t) = \mathcal T \exp\left(-i\int_0^t ds H(s)\right)$ satisfying
\begin{align}
	&U(t)\ (a\ket{0}+b\ket{1})_{c} \ket{\bar 0}_{C\setminus c} = a \ket{\bar 0}_C + b \ket{\bar 1}_C\label{eq:UGHZ}
\end{align}
at time
\begin{align}
	t(r) \leq K_\alpha\times \begin{cases}
		\log^{\kappa_\alpha} r & \text{if } d< \alpha< 2d,\\
		e^{\gamma\sqrt{\log r}}& \text{if } \alpha = 2d, \text{and }\\
		r^{\alpha-2d} & \text{if }2d<\alpha\leq 2d+1.
	\end{cases} \label{eq:Tbound}
\end{align}
Here, $\gamma = 3\sqrt{d}$, $\kappa_\alpha$, and $K_\alpha$ are constants independent of $t$ and $r$.
Additionally, by reversing the unitary in \cref{eq:UGHZ} to ``concentrate'' the information in $\GHZab$ onto a different site in $C$, we can transfer a quantum state from $c\in C$ to any other site $c'\in C$ in time $2t$.

\section{Optimal protocol}\label{sec:protocol}
The key idea of our protocol (\cref{fig:protocol}) is to recursively build the GHZ-like state in a large hypercube from the GHZ-like states of smaller hypercubes.
For the base case, we note that hypercubes of finite lengths, i.e.\ $r\leq r_0$ for some fixed $r_0$, can always be generated in times that satisfy \cref{eq:Tbound} for some suitably large (but constant) prefactor $K_\alpha$.
Assuming that we can encode information into a GHZ-like state in hypercubes of length $r_1$ in time $t_1$ satisfying \cref{eq:Tbound}, the following subroutine encodes information into a GHZ-like state in an  arbitrary
hypercube $C$ of length $r = m r_1$ containing $c$---the site initially holding the phase information $a,b$. Here $m$ is an $\alpha$-dependent number to be chosen later.

\textbf{Step 1:}
We divide the hypercube $C$ into $m^d$ smaller hypercubes $C_1,\dots,C_{m^d}$, each of length $r_1$.
Without loss of generality, let $C_1$ be the hypercube that contains $c$.
Let $V = r_1^d$ be the number of sites in each $C_j$.
In this step, we simultaneously encode $a,b$ into $\GHZab_{C_1}$ and prepare $\GHZ_{C_j}$ for all $j=2,\dots,m^d$, which, by our assumption, takes time
\begin{align}
	t_1  \leq K_\alpha\times\begin{cases}
		\log^{\kappa_\alpha} r_1 & \text{if } d< \alpha < 2d,\\
		e^{\gamma \sqrt{\log r_1}} &\text{if } \alpha = 2d, \text{and} \\
		r_1^{\alpha-2d} & \text{if }2d< \alpha\leq 2d+1.
	\end{cases} \label{eq:t1bound}
\end{align}
By the end of this step, the hypercube $C$ is in the state
\begin{align}
	(a\ket{\bar 0} + b\ket{\bar 1})_{C_1}\bigotimes_{j=2}^{m^d} \frac{\ket{\bar 0}_{C_j} + \ket{\bar 1}_{C_j}}{\sqrt{2}}.
\end{align}

\textbf{Step 2:}
Next, we apply the following Hamiltonian to the hypercube $C$:
\begin{align}
	H_2 = \frac{1}{(mr_1\sqrt{d})^\alpha} \sum_{j = 2}^{m^d} \sum_{\mu\in C_1} \sum_{\nu\in C_j} \ket{1}\bra{1}_\mu\otimes \ket{1}\bra{1}_\nu. \label{eq:cPhase}
\end{align}
This Hamiltonian effectively generates the so-called controlled-PHASE gate between the hypercubes, with $C_1$ being the control hypercube and $C_2,\dots, C_{m^d}$ being the target hypercubes.
We choose the interactions between qubits in \cref{eq:cPhase} to be identical for simplicity. 
If the interactions were to vary between qubits, we would simply turn off the interaction between $C_1$ and $C_j$ once the total phase accumulated by $C_j$ reaches $\pi$
~\footnote{Because only the total accumulated phase matters in choosing the evolution time, we also expect the protocol to be robust against experimental errors such as uncertainties in the positions of individual particles:
If the position of each particle is known up to a precision $\epsilon\ll 1$, the total worst-case error in the accumulated phase scales as $t({r_1^{2d}}/{r^\alpha})\times (\epsilon/r)$, with $r_1$ being the length of each hypercubes and $r$ being the minimum distance between them. The result is a relative phase error proportional to $\epsilon/r$, which becomes smaller and smaller as the distance between the hypercubes increases. 
Moreover, we expect the relative error to be even smaller in the commonly occurring situation when uncertainties in the positions are uncorrelated between different particles
}.  
The prefactor $1/(mr_1\sqrt{d})^\alpha$ ensures that this Hamiltonian satisfies the condition of a power-law interacting Hamiltonian.
It is straightforward to verify that, under this evolution, the state of the hypercube $C$ rotates to
\begin{align}
	a
	\ket{\bar 0}_{C_1}\bigotimes_{j=2}^{m^d}\frac{\ket{\bar 0}_{C_j} + \ket{\bar 1}_{C_j}}{\sqrt 2}+
	b \ket{\bar 1}_{C_1}\bigotimes_{j=2}^{m^d}\frac{\ket{\bar 0}_{C_j} - \ket{\bar 1}_{C_j}}{\sqrt{2}}
\end{align}
after time $t_2 = \pi d^{\alpha/2}(mr_1)^\alpha/{V^2}$.

To obtain the desired state $\GHZab_C$, it remains to apply a Hadamard gate on the effective qubit $\{\ket{\bar 0}_{C_j},\ket{\bar 1}_{C_j}\}$ for $j = 2, \dots, m^d$. We do this in the following three steps by first concentrating the information stored in hypercube $C_j$ onto a single site $c_j \in C_j$  (Step 3), then applying a Hadamard gate on $c_j$ (Step 4), and then unfolding the information back onto the full hypercube $C_j$ (Step 5).

\textbf{Step 3:}
By our assumption, for each hypercube $C_j$ ($j = 2,\dots, m^d$) and given a designated site $c_j\in C_j$, there exists a (time-dependent) Hamiltonian $H_j$ that generates a unitary $U_j$ such that
\begin{align}
	(\psi_0\ket{0}+\psi_1\ket{1})_{c_j} \ket{\bar 0}_{C_j \setminus c_j}
	\xrightarrow{U_j} \psi_0 \ket{\bar 0}_{C_j} + \psi_1\ket{\bar 1}_{C_j}
\end{align}
for all complex coefficients $\psi_0$ and $\psi_1$, in time $t_1$ satisfying \cref{eq:t1bound}.
By linearity, this property applies even if $C_j$ is entangled with other hypercubes.
Consequently, backward time evolution under $H_j$ generates $U_j^\dag$, which ``undoes'' the GHZ-like state of the $j$th hypercube:
\begin{align}
	  \psi_0 \ket{\bar 0}_{C_j} + \psi_1 \ket{\bar 1}_{C_j}
	\xrightarrow{U_j^\dagger} (\psi_0 \ket{0}+\psi_1\ket{1})_{c_j} \ket{\bar 0}_{C_j \setminus c_j}
\end{align}
for any $\psi_0,\psi_1$.
In this step, we simultaneously apply $U_j^\dagger$ to $C_j$ for all $j = 2,\dots,m^d$.
These unitaries rotate the state of $C$ to
\begin{align}
	 a\ket{\bar 0}_{C_1} \bigotimes_{j=2}^{m^d} \ket{+}_{c_j}\ket{\bar 0}_{C_j \setminus c_j}
	 + b\ket{\bar 1}_{C_1} \bigotimes_{j=2}^{m^d} \ket{-}_{c_j} \ket{\bar 0}_{C_j \setminus c_j},
\end{align}
where $\ket{\pm} = (\ket{0}\pm\ket{1})/{\sqrt{2}}$.

\textbf{Step 4:} We then apply a Hadamard gate, i.e.
\begin{align}
	\frac{1}{\sqrt{2}}
 	\begin{pmatrix}
 	 		1 & 1\\
 	 		1 &-1
 	 	\end{pmatrix},
 \end{align} to the site $c_j$ of each hypercubes $C_j$, $j = 2,\dots,m^d$.
These Hadamard gates can be implemented arbitrarily fast since we do not assume any constraints on the single-site terms of the Hamiltonian. The state of $C$ by the end of this step is
\begin{align}
	  a\ket{\bar 0}_{C_1} \bigotimes_{j=2}^{m^d} \ket{0}_{c_j} \ket{\bar 0}_{C_j \setminus c_j}
	 + b\ket{\bar 1}_{C_1} \bigotimes_{j=2}^{m^d} \ket{1}_{c_j} \ket{\bar 0}_{C_j \setminus c_j}.
\end{align}

\textbf{Step 5:} Finally, we apply $U_j$ again to each hypercube $C_j$ ($j=2,\dots, m^d$) to obtain the desired GHZ-like state:
\begin{align}
	 a\ket{\bar 0}_{C_1} \bigotimes_{j=2}^{m^d}  \ket{\bar 0}_{C_j }
	  +b\ket{\bar 1}_{C_1} \bigotimes_{j=2}^{m^d} \ket{\bar 1}_{C_j} = \GHZab_C.
\end{align}

At the end of this routine, we have implemented the unitary satisfying \cref{eq:UGHZ} in time
\begin{align}
	t = 3t_1 + t_2
	= 3t_1 + \pi d^{\alpha/2} m^\alpha r_1^{\alpha-2d}. \label{eq:Tsum}
\end{align}
We now consider three cases corresponding to different ranges of $\alpha$ and show that if $t_1(r_1)$ satisfies \cref{eq:Tbound}, then $t(r)$ also satisfies \cref{eq:Tbound}.

For $\alpha \in (2d,2d+1]$, we have $t_1 \leq K_\alpha r_1^{\alpha-2d}$.
Choosing $m>1$ to be a constant integer, we have
\begin{align}
	 t
	 &\leq  \bigg(\frac{3K_\alpha}{m^{\alpha-2d}}+\pi d^{\alpha/2} m^{2d}\bigg) (mr_1)^{\alpha-2d} \leq K_\alpha  r^{\alpha-2d}, \label{eq:Tbound1}
\end{align}
where we require $m > 3^{1/(\alpha-2d)}$ and choose
\begin{align}
	K_\alpha \geq \frac{\pi d^{\alpha/2} m^{2d}}{1 - \frac{3}{m^{\alpha-2d}}} = \frac{\pi d^{\alpha/2} m^\alpha}{m^{\alpha-2d}-3}. \label{eq:K1}
\end{align}

For $\alpha \in (d,2d)$, we choose $m$ to scale with $r_1$ such that $r_1^{\lambda - 1} < m \leq 2 r_1^{\lambda - 1}$ where $\lambda = 2d/\alpha$.
The length of the larger cube $C$ is then $r = mr_1 > r_1^{\lambda}$ and, therefore, the total time is
\begin{align}
	t &\leq 3K_\alpha \log^{\kappa_\alpha} r_1 +\pi (2\sqrt d)^\alpha r_1^{(\lambda-1)\alpha + \alpha-2d}\\
	&\leq \frac{4K_\alpha}{\lambda^{\kappa_\alpha}} \log^{\kappa_\alpha} \left(r_1^\lambda\right)
	\leq K_\alpha \log^{\kappa_\alpha} r,\label{eq:Tbound2}
\end{align}
where we choose $\kappa_\alpha = \log 4 / \log(2d/\alpha)$ and assume $K_\alpha \log^{\kappa_\alpha}r_1 \geq \pi(2\sqrt{d})^\alpha$ to simplify the expression.
We note that the factor $\log 4$ in the definition of $\kappa_\alpha$ can be made arbitrarily close to $\log 3$ by increasing $K_\alpha$.

Finally, for $\alpha = 2d$, we choose $m$ such that $\exp(\frac{\gamma}{2d}\sqrt{\log r_1})\leq m \leq 2\exp(\frac{\gamma}{2d}\sqrt{\log r_1})$, where $\gamma = 3\sqrt{d}$.
Substituting $t_1\leq K_\alpha \exp(\gamma\sqrt{\log{r_1}})$ into \cref{eq:Tsum}, we have
\begin{align}
	t \leq \left(3K_\alpha + 2^{\alpha}\pi d^{\alpha/2}\right) e^{\gamma\sqrt{\log r_1}}.
\end{align}
Assuming $r_1 \geq \exp(8/d)$, it is straightforward to prove that $\gamma\sqrt{\log r_1} \leq \gamma\sqrt{\log(mr_1)} - 2$.
Applying this condition on the above inequality, we have
\begin{align}
	t \leq  \frac{1}{e^2} \left(3K_\alpha + 2^{\alpha}\pi d^{\alpha/2}\right) e^{\gamma\sqrt{\log r}}
	\leq K_\alpha e^{\gamma \sqrt{\log r}},\label{eq:Tbound3}
\end{align}
where $r = mr_1$ is the length of the resulting GHZ-like state and we chose $K_\alpha \geq 2^\alpha \pi d^{\alpha/2}/(e^2-3)$.
\Cref{eq:Tbound1,eq:Tbound2,eq:Tbound3} prove that $t$ satisfies \cref{eq:Tbound}.
Repeatedly applying this routine yields larger and larger GHZ-like states.

Before discussing the implications of our protocol, we would like to explain intuitively the main sources of its improvement relative to existing protocols.
In our protocol, we simultaneously encode the information into the GHZ-like state over $C_1$ and create the symmetric GHZ states over other multiqubit subsystems $C_2,\dots,C_{m^{d}}$.
As a result, the implementation of the controlled operations in step 2 (\cref{fig:protocol}) is enhanced quadratically by the volume of each subsystems.
In contrast, the protocol in Ref.~\cite{Eldredge17} applies controlled operations between a large subsystem and individual remaining sites of the system, resulting in the implementation time scaling only linearly with the volume of the subsystem.

On the other hand, while the state transfer protocol in Refs.~\cite{tranHierarchyLinearLight2020a,kuwaharaStrictlyLinearLight2020} also applies controlled operations between large subsystems and is, therefore, sped up quadratically by the subsystem volume, it only uses qubits in small neighborhoods around the source and the target of the transfer.
In our protocol, we maximize the size of the resulting GHZ-like state at the end of each iteration by allowing $m$ to depend on $\alpha$ and on the size of the existing GHZ-like states.
When we use the protocol for state transfer, this strategy results in most of the qubits between the source and the target sites participating in the transfer, significantly speeding up the protocol.

\begin{table*}[t]
	\resizebox{2.05\columnwidth}{!}{%
	\begin{tabular}{llll}
		\toprule
		Tasks&Known light cones&Previous best protocols&Our protocol\\
		\midrule
		\begin{tabular}{l}
		 	Encoding into\\
		 	a GHZ-like
		 	state
		 \end{tabular} &
		 $ t\gtrsim
		 \begin{cases}
		 	\log r & \alpha \in (d,2d]~\mysupcite{HK}\\
		 	r^{\frac{\alpha-2d}{\alpha-d}} & \alpha \in (2d,2d+1), d>1~\mysupcite{tranLocalityDigitalQuantum2019a}\\
		 	r^{\alpha-2} & \alpha \in (2,3), d = 1~\mysupcite{chenFiniteSpeedQuantum2019}
		 \end{cases}
		$
		 &
		 $t \sim
		 \begin{cases}
		 	r^{\alpha-d} & \alpha \in (d,d+1)~\mysupcite{Eldredge17}\\
		 	r & \alpha \in [d+1,2d+1)
		 \end{cases}
		$
		 &
		 $t \sim
		 \begin{cases}
		 	\polylog(r) &\alpha \in (d,2d)\\
		 	e^{\gamma \sqrt{\log r}} &\alpha = 2d\\
		 	r^{\alpha-2d} & \alpha \in (2d,2d+1)
		 \end{cases}
		$
		 \\\addlinespace[0.75em]
		 %%%%%%%%%%%%%%%%%%%%%%%%%%%%%%%%
		 % \midrule
		 \begin{tabular}{l}
		 	Preparing  a known\\
		 	GHZ-like
		 	state
		 \end{tabular} &
		 $ t\gtrsim
		 \begin{cases}
		 	\log r & \alpha \in (d,2d]~\mysupcite{HK,Bravyi06}\\
		 	r^{\frac{\alpha-2d}{\alpha-d+1}} & \alpha \in (2d,2d+1)~\mysupcite{GongFF,tranLocalityDigitalQuantum2019a}
		 \end{cases}
		$
		 &
		 \begin{tabular}{l}
		 	Same as encoding\\
		 	into a GHZ-like state
		 \end{tabular}
		 &
		 Same as above
		 \\\addlinespace[0.75em]
		 %%%%%%%%%%%%%%%%%%%%%%%%%%%%%%%%
		 % \midrule
		 \begin{tabular}{l}
		 	State
		 	transfer
		 \end{tabular} &
		 \begin{tabular}{l}
		 	Same as encoding \\
		 	into a GHZ-like state
		 \end{tabular}
		 &
		 $t \sim
		 \begin{cases}
		 	r^{\frac{\alpha(\alpha-d)}{\alpha+d}}& \alpha \in (d,d+1)~\mysupcite{tranHierarchyLinearLight2020a}\\
		 	r^{\frac{\alpha}{2d+1}} & \alpha \in [d+1,2d+1)~\mysupcite{tranHierarchyLinearLight2020a,kuwaharaStrictlyLinearLight2020}
		 \end{cases}
		$
		 &
		 Same as above
		 \\\addlinespace[0.75em]
		 %%%%%%%%%%%%%%%%%%%%%%%%%%%%%%%%
		 % \midrule
		 \begin{tabular}{l}
		 	State transfer\\
		 	(no initialization)
		 \end{tabular} &
		 $ t\gtrsim
		 \begin{cases}
		 	r^{\frac{2\alpha-2d}{2\alpha-d+1}} & \alpha \in (d,2d]~\mysupcite{kuwaharaPolynomialGrowthOutoftimeorder2020a}\\
		 	r^{\frac{\alpha-2d}{\alpha-d}} & \alpha \in (2d,2d+1)~\mysupcite{tranLocalityDigitalQuantum2019a}
		 	\\
		 	r^{\alpha-\frac{3}{2}} & \alpha \in (2,\frac52], d = 1~\mysupcite{tranHierarchyLinearLight2020a}
		 	\\
		 	r & \alpha \in (\frac52,3), d = 1~\mysupcite{tranHierarchyLinearLight2020a}
		 \end{cases}
		$
		 &
		 $t\sim r$ $\ \forall \alpha \in (d,2d+1)$
		 &
		 Not applicable
		 \\\addlinespace[0.5em]
		 %%%%%%%%%%%%%%%%%%%%%%%%%%%%%%%%%%%%%%%%%%%%%%%%%%%%%%%%%%%%%%%%%
		\botrule
	\end{tabular}
	}
	\caption{A summary of known bounds and protocols in the regime $\alpha \in (d,2d+1)$ for several information-propagation tasks: encoding an unknown qubit state into a GHZ-like state (row 1), preparing a known GHZ-like state (row 2), state transfer assuming we can initialize intermediate qubits (row 3), and state transfer given intermediate qubits in arbitrary states (i.e.\ so-called universal state transfer \cite{tranHierarchyLinearLight2020a}, row 4).
	The tasks of encoding information into GHZ-like states and quantum state transfer with initialization are constrained by the Lieb-Robinson bounds.
	Preparing a known GHZ-like state, being potentially easier than encoding unknown information into GHZ-like states, is---at least at present---sometimes bounded by a weaker light cone~\cite{GongFF,tranLocalityDigitalQuantum2019a}.
	On the other hand, state transfer given intermediate qubits in arbitrary states (i.e.\ universal state transfer) is more difficult than state transfer with initialized intermediate qubits and is bounded by the more stringent Frobenius light cone~\cite{tranHierarchyLinearLight2020a}.
	The bounds on encoding information into GHZ-like states (except Ref.~\cite{chenFiniteSpeedQuantum2019}) also apply to general $k$-body interactions.
	All listed bounds also hold not just for qubits, but for all finite-level systems.
	For $d<\alpha\leq 2d$, our protocol saturates (up to subpolynomial corrections) the known bounds, thus proving the optimality of both the protocol and the bounds.
	For $2d<\alpha<2d+1$, our protocol also saturates the bounds on encoding information into GHZ-like states and state transfer in $d=1$ dimension (thus again proving the optimality of both the protocol and the bounds) and suggests what the tightest possible Lieb-Robinson light cone might be for $d>1$ dimensions.
	}
	\label{tab:compare}
\end{table*}

\section{Discussion}\label{sec:discussion}
We now discuss the performance and the implications of our protocol (summarized in \cref{tab:compare}).
First, our protocol allows for encoding an unknown qubit into a multiqubit
GHZ-like state and, subsequently, performing state transfer at unprecedented speeds.
For $d<\alpha< 2d$, which applies, for example, to dipole-dipole interactions ($\alpha = 3$) in two dimensions and to the effective interactions between trapped ions ($\alpha\in [0,3]$) in one and two dimensions, our protocol encodes information into GHZ-like states and transfers information in polylogarithmic time, exponentially faster than protocols available in the literature.
Even for the seemingly weakly long-range interactions with $\alpha = 2d$, such as van der Waals interactions between Rydberg atoms ($\alpha = 6$) in three dimensions, our protocol still takes only subpolynomial time to entangle an entire system and to transfer a quantum state.
When applied to the preparation of GHZ states, these speedups enable potential improvements to quantum sensors built from nitrogen-vacancy centers~\cite{mazeNanoscaleMagneticSensing2008,doldeSensingElectricFields2011a}, Rydberg atoms~\cite{sedlacekMicrowaveElectrometryRydberg2012,wadeRealTimeNearFieldTerahertz2017}, and polar molecules~\cite{carrColdUltracoldMolecules2009}, as well as to atomic clocks based on trapped ions~\cite{andreStabilityAtomicClocks2004}.

\emph{Optimal quantum information processing.---} The optimality of our protocol for $\alpha \in (1,3)$ in one dimension and $\alpha \in (d,2d]$ in $d>1$ dimensions also lays the foundation for optimal quantum information processing in power-law interacting systems~\cite{brittonEngineeredTwodimensionalIsing2012,kimQuantumSimulationTransverse2011}.
Using quantum state transfer between auxiliary qubits and encoding qubits into large GHZ-like states as  subroutines, our protocol leads to optimal implementations of quantum gates between distant qubits in large quantum computers.
In particular, the faster encoding of information into a GHZ-like state of ancillary qubits speeds up \cite{guoImplementingFastUnbounded2020} the implementation of the quantum fanout---a powerful multiqubit quantum gate~\cite{hoyerQuantumFanoutPowerful2005}. At the same time, the faster state transfer speeds up~\cite{Eldredge17} the constructions of multiscale entanglement renormalization ansatz (MERA) states, commonly used to represent highly entangled---including topologically ordered
\cite{aguadoEntanglementRenormalizationTopological2008}---states~\cite{vidalEntanglementRenormalization2007,vidalClassQuantumManyBody2008,giovannettiQuantumMultiscaleEntanglement2008}.
Specifically, we can implement a fanout gate %(by encoding information into a GHZ-like state of ancillary qubits~
\cite{guoImplementingFastUnbounded2020} on qubits in a hypercube of volume $n$ and prepare a MERA state~\cite{Eldredge17} on these qubits in time $t\sim \polylog(n)$ for $\alpha \in (d,2d)$, $t\sim e^{\frac{\gamma}{\sqrt{d}} \sqrt{\log n}}$ for $\alpha = 2d$---which are both exponential speedups compared to the previous best---and $t\sim n^{(\alpha-2d)/d}$ for $\alpha\in (2d,2d+1)$.
The optimality of these operations is again guaranteed (up to subpolynomial corrections) by the matching lower limits imposed by the Lieb-Robinson bounds~\cite{Eldredge17,guoImplementingFastUnbounded2020}.

In practice, using single-site Hamiltonians to implement the echoing technique of Ref.\,\cite{Eldredge17}, the controlled-PHASE gate in step 2 of our protocol can be realized starting from time-independent power-law interactions between all sites of the system. The protocol therefore does not require explicit time-dependent control of individual two-qubit Hamiltonians, making it appealing for implementation on available experimental platforms.
However, because the diameter of the GHZ-like state increases by more than twofold in every iteration of the protocol, the scaling in \cref{eq:Tbound} may only be observed in large systems.

\emph{Information-propagation speed limits.---} Conceptually, since our protocol saturates (up to subpolynomial corrections) the Lieb-Robinson bounds for $d<\alpha\leq 2d$ for all $d$ and, additionally, $2<\alpha<3$ for $d = 1$, we demonstrate, for the first time, the tightness of these fundamental bounds in these regimes.
In particular, the subpolynomial entanglement time for $\alpha\leq 2d$ disproves the conjecture in Ref.~\cite{tranLocalityHeatingPeriodically2019}, where a gap in the provable heating times of periodically driven, power-law interacting systems had suggested the existence of a tighter Lieb-Robinson bound with an algebraic light cone in this regime of $\alpha$.
It would be interesting to determine what could have resulted in this gap in our understanding of the heating time.
Additionally, for $2d<\alpha<2d+1$, our protocol suggests that $t\gtrsim r^{\alpha-2d}$ is the tightest possible light cone, providing strong evidence for the conjectured generalization of the Lieb-Robinson bound in Ref.~\cite{chenFiniteSpeedQuantum2019} to $d>1$.

Since the best known generalizations of these bounds to $k$-body, power-law interacting Hamiltonians---those described by $H = \sum_{X} h_{X}$, where the sum is over all subsets $X\subset \Lambda$ of at most $k$ sites and $\sum_{X \ni {i,j}}\norm{h_X}\leq 1/\dist(i,j)^\alpha$ for all $i\neq j$---have the same scaling as the best known 2-body bounds when $d<\alpha\leq 2d$~\cite{HK} (see also \cref{tab:compare}), the scaling of our 2-body protocol is also optimal even if one allows for $k$-body interactions. In other words, in this regime of $\alpha$, allowing for $k$-body interactions cannot enable a qualitative speedup relative to 2-body interactions.

Our protocol also generalizes straightforwardly from two-level to arbitrary finite-level systems.
Given a $q$-level system at each site of the lattice,
we can unitarily encode an arbitrary state $\ket{\psi}_c = \sum_{\ell=0}^{q-1} a_\ell \ket{\ell}$ of site $c\in C$, where $a_\ell$ are complex coefficients and $C$ is a hypercube of linear size $r$, into a multi-qudit state
\begin{align}
	\ket{\psi}_c \ket{\bar 0}_{C\setminus c}\rightarrow \sum_{\ell=0}^{q-1} a_\ell \ket{\bar \ell}_C
\end{align}
in time $t(r)$ satisfying \cref{eq:Tbound}.
This can be done by replacing the Hamiltonian in \cref{eq:cPhase} with
\begin{align}
	\frac{1}{(mr_1\sqrt{d})^\alpha} \sum_{j = 2}^{m^d} \sum_{\mu\in C_1} \sum_{\nu\in C_j}\sum_{\ell,\ell' = 0}^{q-1} \ell\ell' \ket{\ell}\bra{\ell}_\mu\otimes \ket{\ell'}\bra{\ell'}_\nu \label{eq:cPhasequdit}
\end{align}
and replacing the single-qubit Hadamard gate in step 4 by a $q$-by-$q$ discrete Fourier transform matrix.
Since the Lieb-Robinson bounds have the same light cones for any finite-level systems, our protocol also saturates these bounds for $\alpha \in (1,3)$ in one dimension and $\alpha \in (d,2d]$ in $d>1$ dimensions.

In our protocol, we assume that $a\ket{0} + b\ket{1}$ is a possibly unknown state.
Encoding such a state into the GHZ-like state is at least as hard as generating a GHZ-like state with known coefficients $a,b$.
In fact, the latter task is not known to be sufficient for state transfer and, therefore, is not \emph{directly} constrained by the Lieb-Robinson bounds.
Instead, one often indirectly obtains a speed limit for this task by applying the Lieb-Robinson bounds on the growth of two-point connected correlators~\cite{Bravyi06,HK,tranHierarchyLinearLight2020a}.
Consequently, the task of generating a known GHZ-like state could potentially be constrained by a weaker light cone than that of encoding an unknown qubit state into a GHZ-like state (see~\cref{tab:compare}).
Nevertheless, our protocol for encoding into a GHZ-like state saturates (up to subpolynomial corrections) the bound $t \gtrsim \log r$~\cite{Bravyi06,HK} on the growth of connected correlators when $d<\alpha\leq 2d$, implying that knowing the coefficients $a,b$ does not speed up the preparation of the GHZ-like state in this regime.
It remains an interesting open question whether the same statement holds for $\alpha \in (2d,2d+1)$.

We also note that our protocol violates the so-called Frobenius light cone, initially derived in Ref.~\cite{tranHierarchyLinearLight2020a} for $\alpha > 3/2$ in one dimension as part of a hierarchy of speed limits for different types of information propagation in long-range interacting systems and later extended to regimes of smaller $\alpha$ in Ref.~\cite{kuwaharaPolynomialGrowthOutoftimeorder2020a}.
The Frobenius bound, which considers information propagation from the operator-spreading perspective, constrains information-propagation tasks that are more demanding than the tasks that saturate the  Lieb-Robinson bound, and therefore has a more stringent light cone.
For example, quantum state transfer given intermediate qubits in arbitrary initial states (i.e.\ universal state transfer) is constrained by the Frobenius light cone, whereas state transfer assuming initialized intermediate qubits is constrained by the Lieb-Robinson bound and can actually violate the Frobenius light cone~\cite{tranHierarchyLinearLight2020a} (see also \cref{tab:compare}).
Determining which of the bounds tightly constrains a given task is still an active area of research.
The protocol in this manuscript proves for the first time that the task of encoding information into GHZ-like state---which is at least as hard as state transfer with initialization---is not constrained by the Frobenius light cone, but is instead tightly constrained (up to subpolynomial corrections) by the Lieb-Robinson bound. 
In particular, when $d<\alpha<2d$, our protocol proves that state transfer with initialization can be implemented exponentially faster than state transfer without initialization, which is constrained by polynomial light cones in this regime~\cite{tranHierarchyLinearLight2020a,kuwaharaPolynomialGrowthOutoftimeorder2020a}.
Furthermore, since our protocol for encoding into a GHZ-like state can also be used to prepare a known GHZ-like state, our protocol also proves for the first time that preparing a known GHZ-like state is not constrained by the Frobenius light cone.

An interesting open question is whether our optimal protocol can be generalized to the regime $0\leq \alpha\leq d$, where there are still substantial gaps between the Lieb-Robinson bounds and achievable protocols~\cite{Storch15,eisertBreakdownQuasilocalityLongRange2013,haukeSpreadCorrelationsLongRange2013a,guoSignalingScramblingStrongly2020a,Eldredge17}.
The bounds suggest that, in addition to the distance, the information-propagation time also depends on the total number of sites on the lattice. 
Consequently, we would expect an optimal protocol to make use of all sites on the lattice, including those that are far from both the source and the target of the propagation.
We consider such a generalization an important future direction.

\emph{Resource lower bound in quantum simulation.---}Our protocol also gives the first known example of a lower bound on the gate count in simulating power-law systems on a quantum computer:
it takes $\Omega({n})$ elementary quantum gates to simulate an $n$-qubit power-law system evolving for time $t \geq t_*$, where  
\begin{align}
	t_*  = \begin{cases}
		\Theta\left(\log^{\kappa_\alpha} n\right) & \text{if } d< \alpha < 2d,\\
		\Theta\left(e^{\gamma \sqrt{(\log n)/d}} \right)&\text{if } \alpha = 2d, \text{and} \\
		\Theta\left(n^{\alpha/d-2} \right) & \text{if }2d< \alpha\leq 2d+1,
	\end{cases} \label{eq:t*-lower-bound}
\end{align}
to constant error.
Indeed, if an algorithm could use fewer than $\Omega(n)$ quantum gates to perform the simulation for times within $t= t_*$ satisfying \cref{eq:t*-lower-bound}, we could use the algorithm to simulate our protocol and prepare an $n$-qubit GHZ state.
However, since an $n$-qubit GHZ state must take $\Omega(n)$ quantum gates to prepare, we would arrive at a contradiction.

Lower bounds on the simulation gate count are valuable benchmarks for the performance of quantum algorithms.
Ref.~\cite{Haah} gives an algorithm for simulating  the time evolution of  finite-range interacting Hamiltonians,
the gate count of which was shown to be optimal via a matching lower bound. 
To date, despite progressively more efficient quantum simulation algorithms~\cite{tranLocalityDigitalQuantum2019a,Childs2019d} in recent literature, no saturable lower bounds are known for power-law systems.
% To simulate an $n$-qubit system evolving for time $t$ to constant error, the gate cost was upper-bounded by $\mathcal{\tilde O}(nt)$, a scaling which was shown to be tight via a matching lower bound. 
% Previous algorithms have yielded progressively better upper bounds to the true gate cost of simulating power-law systems \cite{tranLocalityDigitalQuantum2019a,Childs2019d}.
For example, the analysis of the Suzuki-Trotter product formulas in Ref.~\cite{Childs2019d} results in upper bounds
\begin{equation}
     g_\alpha = \begin{cases}
		\O{n^{2+o(1)}t^{1+o(1)}} & \text{if } d< \alpha \leq 2d,\\
		\O{(nt)^{1+d/(\alpha-d)+o(1)}}  & \text{if }\alpha > 2d,
	\end{cases}
     \label{eq:gatecost_ub}
\end{equation} 
for simulating an $n$-qubit power-law system for time $t$. 
At $t = t_*$ given in \cref{eq:t*-lower-bound}, the corresponding upper bounds reduce to
\begin{align}
	g_\alpha = \begin{cases}
		\O{n^{2+o(1)}} & \text{if } d< \alpha \leq 2d,\\
		\O{n^{\alpha/d+o(1)}}  & \text{if }2d < \alpha \leq 2d+1.
	\end{cases}
     \label{eq:gatecost_ub2}
\end{align}
The gap between this state-of-the-art upper bound and our lower bound $\Omega(n)$ hints at the possibility of a more efficient algorithm for simulating power-law systems.

\begin{acknowledgments}
We thank Chi-Fang Chen, Adam Ehrenberg, Yifan Hong, Zhe-Xuan Gong, Dhruv Devulapalli, Aniruddha Bapat, Eddie Schoute, and Andrew Childs for helpful discussions. MCT, AD, AYG, and AVG acknowledge funding by the DoE ASCR Quantum Testbed Pathfinder program (award No.\ DE-SC0019040), DoE ASCR Accelerated Research in Quantum Computing program (award No.\ DE-SC0020312), AFOSR MURI, AFOSR, NSF PFCQC program, ARO MURI, U.S.\  Department of Energy Award No.\  DE-SC0019449, ARL CDQI, and NSF PFC at JQI.
MCT acknowledges additional support from the Princeton Center for Complex Materials, a MRSEC supported by NSF grant DMR 1420541.
AL was supported by a Research Fellowship from the Alfred P. Sloan Foundation.
\end{acknowledgments}

%%%%%%%%%%%%%%%%%%%%%%%%%%%%%%%%%%%%%%%%%%%%%
% \bibliographystyle{abbrv}
\bibliography{my-bib}

\begin{thebibliography}{61}%
\makeatletter
\providecommand \@ifxundefined [1]{%
 \@ifx{#1\undefined}
}%
\providecommand \@ifnum [1]{%
 \ifnum #1\expandafter \@firstoftwo
 \else \expandafter \@secondoftwo
 \fi
}%
\providecommand \@ifx [1]{%
 \ifx #1\expandafter \@firstoftwo
 \else \expandafter \@secondoftwo
 \fi
}%
\providecommand \natexlab [1]{#1}%
\providecommand \enquote  [1]{``#1''}%
\providecommand \bibnamefont  [1]{#1}%
\providecommand \bibfnamefont [1]{#1}%
\providecommand \citenamefont [1]{#1}%
\providecommand \href@noop [0]{\@secondoftwo}%
\providecommand \href [0]{\begingroup \@sanitize@url \@href}%
\providecommand \@href[1]{\@@startlink{#1}\@@href}%
\providecommand \@@href[1]{\endgroup#1\@@endlink}%
\providecommand \@sanitize@url [0]{\catcode `\\12\catcode `\$12\catcode
  `\&12\catcode `\#12\catcode `\^12\catcode `\_12\catcode `\%12\relax}%
\providecommand \@@startlink[1]{}%
\providecommand \@@endlink[0]{}%
\providecommand \url  [0]{\begingroup\@sanitize@url \@url }%
\providecommand \@url [1]{\endgroup\@href {#1}{\urlprefix }}%
\providecommand \urlprefix  [0]{URL }%
\providecommand \Eprint [0]{\href }%
\providecommand \doibase [0]{https://doi.org/}%
\providecommand \selectlanguage [0]{\@gobble}%
\providecommand \bibinfo  [0]{\@secondoftwo}%
\providecommand \bibfield  [0]{\@secondoftwo}%
\providecommand \translation [1]{[#1]}%
\providecommand \BibitemOpen [0]{}%
\providecommand \bibitemStop [0]{}%
\providecommand \bibitemNoStop [0]{.\EOS\space}%
\providecommand \EOS [0]{\spacefactor3000\relax}%
\providecommand \BibitemShut  [1]{\csname bibitem#1\endcsname}%
\let\auto@bib@innerbib\@empty
%</preamble>
\bibitem [{\citenamefont {Wineland}\ \emph {et~al.}(1992)\citenamefont
  {Wineland}, \citenamefont {Bollinger}, \citenamefont {Itano}, \citenamefont
  {Moore},\ and\ \citenamefont {Heinzen}}]{winelandSpinSqueezingReduced1992}%
  \BibitemOpen
  \bibfield  {author} {\bibinfo {author} {\bibfnamefont {D.~J.}\ \bibnamefont
  {Wineland}}, \bibinfo {author} {\bibfnamefont {J.~J.}\ \bibnamefont
  {Bollinger}}, \bibinfo {author} {\bibfnamefont {W.~M.}\ \bibnamefont
  {Itano}}, \bibinfo {author} {\bibfnamefont {F.~L.}\ \bibnamefont {Moore}},\
  and\ \bibinfo {author} {\bibfnamefont {D.~J.}\ \bibnamefont {Heinzen}},\
  }\bibfield  {title} {\bibinfo {title} {Spin squeezing and reduced quantum
  noise in spectroscopy},\ }\href {https://doi.org/10.1103/PhysRevA.46.R6797}
  {\bibfield  {journal} {\bibinfo  {journal} {Phys. Rev. A}\ }\textbf {\bibinfo
  {volume} {46}},\ \bibinfo {pages} {R6797} (\bibinfo {year}
  {1992})}\BibitemShut {NoStop}%
\bibitem [{\citenamefont {Bollinger}\ \emph {et~al.}(1996)\citenamefont
  {Bollinger}, \citenamefont {Itano}, \citenamefont {Wineland},\ and\
  \citenamefont {Heinzen}}]{bollingerOptimalFrequencyMeasurements1996a}%
  \BibitemOpen
  \bibfield  {author} {\bibinfo {author} {\bibfnamefont {J.~J.~.}\ \bibnamefont
  {Bollinger}}, \bibinfo {author} {\bibfnamefont {W.~M.}\ \bibnamefont
  {Itano}}, \bibinfo {author} {\bibfnamefont {D.~J.}\ \bibnamefont
  {Wineland}},\ and\ \bibinfo {author} {\bibfnamefont {D.~J.}\ \bibnamefont
  {Heinzen}},\ }\bibfield  {title} {\bibinfo {title} {Optimal frequency
  measurements with maximally correlated states},\ }\href
  {https://doi.org/10.1103/PhysRevA.54.R4649} {\bibfield  {journal} {\bibinfo
  {journal} {Phys. Rev. A}\ }\textbf {\bibinfo {volume} {54}},\ \bibinfo
  {pages} {R4649} (\bibinfo {year} {1996})}\BibitemShut {NoStop}%
\bibitem [{\citenamefont {Wehner}\ \emph {et~al.}(2018)\citenamefont {Wehner},
  \citenamefont {Elkouss},\ and\ \citenamefont
  {Hanson}}]{wehnerQuantumInternetVision2018}%
  \BibitemOpen
  \bibfield  {author} {\bibinfo {author} {\bibfnamefont {S.}~\bibnamefont
  {Wehner}}, \bibinfo {author} {\bibfnamefont {D.}~\bibnamefont {Elkouss}},\
  and\ \bibinfo {author} {\bibfnamefont {R.}~\bibnamefont {Hanson}},\
  }\bibfield  {title} {\bibinfo {title} {Quantum internet: {{A}} vision for the
  road ahead},\ }\href {doi.org/10.1126/science.aam9288} {\bibfield  {journal}
  {\bibinfo  {journal} {Science}\ }\textbf {\bibinfo {volume} {362}} (\bibinfo
  {year} {2018})}\BibitemShut {NoStop}%
\bibitem [{\citenamefont
  {Feynman}(1982)}]{feynmanSimulatingPhysicsComputers1982}%
  \BibitemOpen
  \bibfield  {author} {\bibinfo {author} {\bibfnamefont {R.~P.}\ \bibnamefont
  {Feynman}},\ }\bibfield  {title} {{\selectlanguage {en}\bibinfo {title}
  {Simulating physics with computers}},\ }\href
  {https://doi.org/10.1007/BF02650179} {\bibfield  {journal} {\bibinfo
  {journal} {Int. J. Theor. Phys.}\ }\textbf {\bibinfo {volume} {21}},\
  \bibinfo {pages} {467} (\bibinfo {year} {1982})}\BibitemShut {NoStop}%
\bibitem [{\citenamefont {Shor}(1994)}]{shorAlgorithmsQuantumComputation1994}%
  \BibitemOpen
  \bibfield  {author} {\bibinfo {author} {\bibfnamefont {P.~W.}\ \bibnamefont
  {Shor}},\ }\bibfield  {title} {\bibinfo {title} {Algorithms for quantum
  computation: Discrete logarithms and factoring},\ }in\ \href
  {https://doi.org/10.1109/SFCS.1994.365700} {\emph {\bibinfo {booktitle}
  {Proceedings 35th {{Annual Symposium}} on {{Foundations}} of {{Computer
  Science}}}}}\ (\bibinfo {year} {1994})\ pp.\ \bibinfo {pages}
  {124--134}\BibitemShut {NoStop}%
\bibitem [{\citenamefont {Pham}\ and\ \citenamefont {Svore}(2013)}]{pham13}%
  \BibitemOpen
  \bibfield  {author} {\bibinfo {author} {\bibfnamefont {P.}~\bibnamefont
  {Pham}}\ and\ \bibinfo {author} {\bibfnamefont {K.~M.}\ \bibnamefont
  {Svore}},\ }\bibfield  {title} {\bibinfo {title} {A 2d nearest-neighbor
  quantum architecture for factoring in polylogarithmic depth},\ }\href
  {https://arxiv.org/abs/1207.6655} {\bibfield  {journal} {\bibinfo  {journal}
  {Quantum Info. Comput.}\ }\textbf {\bibinfo {volume} {13}},\ \bibinfo {pages}
  {937} (\bibinfo {year} {2013})}\BibitemShut {NoStop}%
\bibitem [{\citenamefont {Guo}\ \emph {et~al.}(2020{\natexlab{a}})\citenamefont
  {Guo}, \citenamefont {Deshpande}, \citenamefont {Chu}, \citenamefont
  {Eldredge}, \citenamefont {Bienias}, \citenamefont {Devulapalli},
  \citenamefont {Su}, \citenamefont {Childs},\ and\ \citenamefont
  {Gorshkov}}]{guoImplementingFastUnbounded2020}%
  \BibitemOpen
  \bibfield  {author} {\bibinfo {author} {\bibfnamefont {A.~Y.}\ \bibnamefont
  {Guo}}, \bibinfo {author} {\bibfnamefont {A.}~\bibnamefont {Deshpande}},
  \bibinfo {author} {\bibfnamefont {S.-K.}\ \bibnamefont {Chu}}, \bibinfo
  {author} {\bibfnamefont {Z.}~\bibnamefont {Eldredge}}, \bibinfo {author}
  {\bibfnamefont {P.}~\bibnamefont {Bienias}}, \bibinfo {author} {\bibfnamefont
  {D.}~\bibnamefont {Devulapalli}}, \bibinfo {author} {\bibfnamefont
  {Y.}~\bibnamefont {Su}}, \bibinfo {author} {\bibfnamefont {A.~M.}\
  \bibnamefont {Childs}},\ and\ \bibinfo {author} {\bibfnamefont {A.~V.}\
  \bibnamefont {Gorshkov}},\ }\bibfield  {title} {\bibinfo {title}
  {Implementing a {{Fast Unbounded Quantum Fanout Gate Using Power}}-{{Law
  Interactions}}},\ }\href {http://arxiv.org/abs/2007.00662} {\bibfield
  {journal} {\bibinfo  {journal} {arXiv:2007.00662 [quant-ph]}\ } (\bibinfo
  {year} {2020}{\natexlab{a}})}\BibitemShut {NoStop}%
\bibitem [{\citenamefont {Christandl}\ and\ \citenamefont
  {Wehner}(2005)}]{christandl05}%
  \BibitemOpen
  \bibfield  {author} {\bibinfo {author} {\bibfnamefont {M.}~\bibnamefont
  {Christandl}}\ and\ \bibinfo {author} {\bibfnamefont {S.}~\bibnamefont
  {Wehner}},\ }\bibfield  {title} {\bibinfo {title} {Quantum anonymous
  transmissions},\ }in\ \href {https://arxiv.org/abs/quant-ph/0409201} {\emph
  {\bibinfo {booktitle} {Advances in Cryptology - ASIACRYPT 2005}}},\ \bibinfo
  {editor} {edited by\ \bibinfo {editor} {\bibfnamefont {B.}~\bibnamefont
  {Roy}}}\ (\bibinfo  {publisher} {Springer Berlin Heidelberg},\ \bibinfo
  {address} {Berlin, Heidelberg},\ \bibinfo {year} {2005})\ pp.\ \bibinfo
  {pages} {217--235}\BibitemShut {NoStop}%
\bibitem [{\citenamefont {Brassard}\ \emph {et~al.}(2009)\citenamefont
  {Brassard}, \citenamefont {Broadbent}, \citenamefont {Fitzsimons},
  \citenamefont {Gambs},\ and\ \citenamefont {Tapp}}]{brassard09}%
  \BibitemOpen
  \bibfield  {author} {\bibinfo {author} {\bibfnamefont {G.}~\bibnamefont
  {Brassard}}, \bibinfo {author} {\bibfnamefont {A.}~\bibnamefont {Broadbent}},
  \bibinfo {author} {\bibfnamefont {J.}~\bibnamefont {Fitzsimons}}, \bibinfo
  {author} {\bibfnamefont {S.}~\bibnamefont {Gambs}},\ and\ \bibinfo {author}
  {\bibfnamefont {A.}~\bibnamefont {Tapp}},\ }\bibfield  {title} {\bibinfo
  {title} {Anonymous quantum communication},\ }in\ \href
  {https://arxiv.org/abs/0706.2356} {\emph {\bibinfo {booktitle} {Information
  Theoretic Security}}},\ \bibinfo {editor} {edited by\ \bibinfo {editor}
  {\bibfnamefont {Y.}~\bibnamefont {Desmedt}}}\ (\bibinfo  {publisher}
  {Springer Berlin Heidelberg},\ \bibinfo {address} {Berlin, Heidelberg},\
  \bibinfo {year} {2009})\ pp.\ \bibinfo {pages} {181--182}\BibitemShut
  {NoStop}%
\bibitem [{\citenamefont {Hillery}\ \emph {et~al.}(1999)\citenamefont
  {Hillery}, \citenamefont {Bu{\v z}ek},\ and\ \citenamefont
  {Berthiaume}}]{hilleryQuantumSecretSharing1999}%
  \BibitemOpen
  \bibfield  {author} {\bibinfo {author} {\bibfnamefont {M.}~\bibnamefont
  {Hillery}}, \bibinfo {author} {\bibfnamefont {V.}~\bibnamefont {Bu{\v
  z}ek}},\ and\ \bibinfo {author} {\bibfnamefont {A.}~\bibnamefont
  {Berthiaume}},\ }\bibfield  {title} {\bibinfo {title} {Quantum secret
  sharing},\ }\href {https://doi.org/10.1103/PhysRevA.59.1829} {\bibfield
  {journal} {\bibinfo  {journal} {Phys. Rev. A}\ }\textbf {\bibinfo {volume}
  {59}},\ \bibinfo {pages} {1829} (\bibinfo {year} {1999})}\BibitemShut
  {NoStop}%
\bibitem [{\citenamefont {Lieb}\ and\ \citenamefont {Robinson}(1972)}]{LR}%
  \BibitemOpen
  \bibfield  {author} {\bibinfo {author} {\bibfnamefont {E.~H.}\ \bibnamefont
  {Lieb}}\ and\ \bibinfo {author} {\bibfnamefont {D.~W.}\ \bibnamefont
  {Robinson}},\ }\bibfield  {title} {\bibinfo {title} {{The Finite Group
  Velocity of Quantum Spin Systems}},\ }\href
  {https://projecteuclid.org:443/euclid.cmp/1103858407} {\bibfield  {journal}
  {\bibinfo  {journal} {Comm. Math. Phys.}\ }\textbf {\bibinfo {volume} {28}},\
  \bibinfo {pages} {251} (\bibinfo {year} {1972})}\BibitemShut {NoStop}%
\bibitem [{\citenamefont {Nachtergaele}\ \emph {et~al.}(2006)\citenamefont
  {Nachtergaele}, \citenamefont {Ogata},\ and\ \citenamefont
  {Sims}}]{NachtergaeleOS2006}%
  \BibitemOpen
  \bibfield  {author} {\bibinfo {author} {\bibfnamefont {B.}~\bibnamefont
  {Nachtergaele}}, \bibinfo {author} {\bibfnamefont {Y.}~\bibnamefont
  {Ogata}},\ and\ \bibinfo {author} {\bibfnamefont {R.}~\bibnamefont {Sims}},\
  }\bibfield  {title} {\bibinfo {title} {{Propagation of Correlations in
  Quantum Lattice Systems}},\ }\href
  {https://doi.org/10.1007/s10955-006-9143-6} {\bibfield  {journal} {\bibinfo
  {journal} {J. Stat. Phys.}\ }\textbf {\bibinfo {volume} {124}},\ \bibinfo
  {pages} {1} (\bibinfo {year} {2006})}\BibitemShut {NoStop}%
\bibitem [{\citenamefont {Nachtergaele}\ and\ \citenamefont
  {Sims}(2006)}]{Nachtergaele2006}%
  \BibitemOpen
  \bibfield  {author} {\bibinfo {author} {\bibfnamefont {B.}~\bibnamefont
  {Nachtergaele}}\ and\ \bibinfo {author} {\bibfnamefont {R.}~\bibnamefont
  {Sims}},\ }\bibfield  {title} {\bibinfo {title} {{Lieb-Robinson Bounds and
  the Exponential Clustering Theorem}},\ }\href
  {https://doi.org/10.1007/s00220-006-1556-1} {\bibfield  {journal} {\bibinfo
  {journal} {Communications in Mathematical Physics}\ }\textbf {\bibinfo
  {volume} {265}},\ \bibinfo {pages} {119} (\bibinfo {year}
  {2006})}\BibitemShut {NoStop}%
\bibitem [{\citenamefont {{Hastings}}\ and\ \citenamefont {{Koma}}(2006)}]{HK}%
  \BibitemOpen
  \bibfield  {author} {\bibinfo {author} {\bibfnamefont {M.~B.}\ \bibnamefont
  {{Hastings}}}\ and\ \bibinfo {author} {\bibfnamefont {T.}~\bibnamefont
  {{Koma}}},\ }\bibfield  {title} {\bibinfo {title} {{Spectral Gap and
  Exponential Decay of Correlations}},\ }\href
  {https://doi.org/10.1007/s00220-006-0030-4} {\bibfield  {journal} {\bibinfo
  {journal} {Comm. Math. Phys.}\ }\textbf {\bibinfo {volume} {265}},\ \bibinfo
  {pages} {781} (\bibinfo {year} {2006})}\BibitemShut {NoStop}%
\bibitem [{\citenamefont {{Gong}}\ \emph {et~al.}(2014)\citenamefont {{Gong}},
  \citenamefont {{Foss-Feig}}, \citenamefont {{Michalakis}},\ and\
  \citenamefont {{Gorshkov}}}]{GongFF}%
  \BibitemOpen
  \bibfield  {author} {\bibinfo {author} {\bibfnamefont {Z.-X.}\ \bibnamefont
  {{Gong}}}, \bibinfo {author} {\bibfnamefont {M.}~\bibnamefont {{Foss-Feig}}},
  \bibinfo {author} {\bibfnamefont {S.}~\bibnamefont {{Michalakis}}},\ and\
  \bibinfo {author} {\bibfnamefont {A.~V.}\ \bibnamefont {{Gorshkov}}},\
  }\bibfield  {title} {\bibinfo {title} {{Persistence of Locality in Systems
  With Power-Law Interactions}},\ }\href
  {https://doi.org/10.1103/PhysRevLett.113.030602} {\bibfield  {journal}
  {\bibinfo  {journal} {Phys. Rev. Lett.}\ }\textbf {\bibinfo {volume} {113}},\
  \bibinfo {eid} {030602} (\bibinfo {year} {2014})}\BibitemShut {NoStop}%
\bibitem [{\citenamefont {Foss-Feig}\ \emph {et~al.}(2015)\citenamefont
  {Foss-Feig}, \citenamefont {Gong}, \citenamefont {Clark},\ and\ \citenamefont
  {Gorshkov}}]{Foss-FeigG}%
  \BibitemOpen
  \bibfield  {author} {\bibinfo {author} {\bibfnamefont {M.}~\bibnamefont
  {Foss-Feig}}, \bibinfo {author} {\bibfnamefont {Z.-X.}\ \bibnamefont {Gong}},
  \bibinfo {author} {\bibfnamefont {C.~W.}\ \bibnamefont {Clark}},\ and\
  \bibinfo {author} {\bibfnamefont {A.~V.}\ \bibnamefont {Gorshkov}},\
  }\bibfield  {title} {\bibinfo {title} {{Nearly Linear Light Cones in
  Long-Range Interacting Quantum Systems}},\ }\href
  {https://doi.org/10.1103/PhysRevLett.114.157201} {\bibfield  {journal}
  {\bibinfo  {journal} {Phys. Rev. Lett.}\ }\textbf {\bibinfo {volume} {114}},\
  \bibinfo {pages} {157201} (\bibinfo {year} {2015})}\BibitemShut {NoStop}%
\bibitem [{\citenamefont {Storch}\ \emph {et~al.}(2015)\citenamefont {Storch},
  \citenamefont {Worm},\ and\ \citenamefont {Kastner}}]{Storch15}%
  \BibitemOpen
  \bibfield  {author} {\bibinfo {author} {\bibfnamefont {D.-M.}\ \bibnamefont
  {Storch}}, \bibinfo {author} {\bibfnamefont {M.~V.~D.}\ \bibnamefont
  {Worm}},\ and\ \bibinfo {author} {\bibfnamefont {M.}~\bibnamefont
  {Kastner}},\ }\bibfield  {title} {\bibinfo {title} {{Interplay of Soundcone
  and Supersonic Propagation in Lattice Models With Power Law Interactions}},\
  }\href {http://stacks.iop.org/1367-2630/17/i=6/a=063021} {\bibfield
  {journal} {\bibinfo  {journal} {New J. Phys.}\ }\textbf {\bibinfo {volume}
  {17}},\ \bibinfo {pages} {063021} (\bibinfo {year} {2015})}\BibitemShut
  {NoStop}%
\bibitem [{\citenamefont {Nachtergaele}\ \emph {et~al.}(2009)\citenamefont
  {Nachtergaele}, \citenamefont {Raz}, \citenamefont {Schlein},\ and\
  \citenamefont {Sims}}]{NRSS09}%
  \BibitemOpen
  \bibfield  {author} {\bibinfo {author} {\bibfnamefont {B.}~\bibnamefont
  {Nachtergaele}}, \bibinfo {author} {\bibfnamefont {H.}~\bibnamefont {Raz}},
  \bibinfo {author} {\bibfnamefont {B.}~\bibnamefont {Schlein}},\ and\ \bibinfo
  {author} {\bibfnamefont {R.}~\bibnamefont {Sims}},\ }\bibfield  {title}
  {\bibinfo {title} {Lieb-robinson bounds for harmonic and anharmonic lattice
  systems},\ }\href {https://doi.org/10.1007/s00220-008-0630-2} {\bibfield
  {journal} {\bibinfo  {journal} {Comm. Math. Phys.}\ }\textbf {\bibinfo
  {volume} {286}},\ \bibinfo {pages} {1073} (\bibinfo {year}
  {2009})}\BibitemShut {NoStop}%
\bibitem [{\citenamefont {Pr\'emont-Schwarz}\ \emph {et~al.}(2010)\citenamefont
  {Pr\'emont-Schwarz}, \citenamefont {Hamma}, \citenamefont {Klich},\ and\
  \citenamefont {Markopoulou-Kalamara}}]{SHKM10}%
  \BibitemOpen
  \bibfield  {author} {\bibinfo {author} {\bibfnamefont {I.}~\bibnamefont
  {Pr\'emont-Schwarz}}, \bibinfo {author} {\bibfnamefont {A.}~\bibnamefont
  {Hamma}}, \bibinfo {author} {\bibfnamefont {I.}~\bibnamefont {Klich}},\ and\
  \bibinfo {author} {\bibfnamefont {F.}~\bibnamefont {Markopoulou-Kalamara}},\
  }\bibfield  {title} {\bibinfo {title} {Lieb-robinson bounds for
  commutator-bounded operators},\ }\href
  {https://doi.org/10.1103/PhysRevA.81.040102} {\bibfield  {journal} {\bibinfo
  {journal} {Phys. Rev. A}\ }\textbf {\bibinfo {volume} {81}},\ \bibinfo
  {pages} {040102} (\bibinfo {year} {2010})}\BibitemShut {NoStop}%
\bibitem [{\citenamefont {Pr\'emont-Schwarz}\ and\ \citenamefont
  {Hnybida}(2010)}]{SH10}%
  \BibitemOpen
  \bibfield  {author} {\bibinfo {author} {\bibfnamefont {I.}~\bibnamefont
  {Pr\'emont-Schwarz}}\ and\ \bibinfo {author} {\bibfnamefont {J.}~\bibnamefont
  {Hnybida}},\ }\bibfield  {title} {\bibinfo {title} {Lieb-robinson bounds on
  the speed of information propagation},\ }\href
  {https://doi.org/10.1103/PhysRevA.81.062107} {\bibfield  {journal} {\bibinfo
  {journal} {Phys. Rev. A}\ }\textbf {\bibinfo {volume} {81}},\ \bibinfo
  {pages} {062107} (\bibinfo {year} {2010})}\BibitemShut {NoStop}%
\bibitem [{\citenamefont {Tran}\ \emph
  {et~al.}(2019{\natexlab{a}})\citenamefont {Tran}, \citenamefont {Guo},
  \citenamefont {Su}, \citenamefont {Garrison}, \citenamefont {Eldredge},
  \citenamefont {{Foss-Feig}}, \citenamefont {Childs},\ and\ \citenamefont
  {Gorshkov}}]{tranLocalityDigitalQuantum2019a}%
  \BibitemOpen
  \bibfield  {author} {\bibinfo {author} {\bibfnamefont {M.~C.}\ \bibnamefont
  {Tran}}, \bibinfo {author} {\bibfnamefont {A.~Y.}\ \bibnamefont {Guo}},
  \bibinfo {author} {\bibfnamefont {Y.}~\bibnamefont {Su}}, \bibinfo {author}
  {\bibfnamefont {J.~R.}\ \bibnamefont {Garrison}}, \bibinfo {author}
  {\bibfnamefont {Z.}~\bibnamefont {Eldredge}}, \bibinfo {author}
  {\bibfnamefont {M.}~\bibnamefont {{Foss-Feig}}}, \bibinfo {author}
  {\bibfnamefont {A.~M.}\ \bibnamefont {Childs}},\ and\ \bibinfo {author}
  {\bibfnamefont {A.~V.}\ \bibnamefont {Gorshkov}},\ }\bibfield  {title}
  {\bibinfo {title} {Locality and {{Digital Quantum Simulation}} of
  {{Power}}-{{Law Interactions}}},\ }\href
  {https://doi.org/10.1103/PhysRevX.9.031006} {\bibfield  {journal} {\bibinfo
  {journal} {Phys. Rev. X}\ }\textbf {\bibinfo {volume} {9}},\ \bibinfo {pages}
  {031006} (\bibinfo {year} {2019}{\natexlab{a}})}\BibitemShut {NoStop}%
\bibitem [{\citenamefont {Else}\ \emph {et~al.}(2020)\citenamefont {Else},
  \citenamefont {Machado}, \citenamefont {Nayak},\ and\ \citenamefont
  {Yao}}]{elseImprovedLiebRobinsonBound2018}%
  \BibitemOpen
  \bibfield  {author} {\bibinfo {author} {\bibfnamefont {D.~V.}\ \bibnamefont
  {Else}}, \bibinfo {author} {\bibfnamefont {F.}~\bibnamefont {Machado}},
  \bibinfo {author} {\bibfnamefont {C.}~\bibnamefont {Nayak}},\ and\ \bibinfo
  {author} {\bibfnamefont {N.~Y.}\ \bibnamefont {Yao}},\ }\bibfield  {title}
  {\bibinfo {title} {Improved lieb-robinson bound for many-body hamiltonians
  with power-law interactions},\ }\href
  {https://doi.org/10.1103/PhysRevA.101.022333} {\bibfield  {journal} {\bibinfo
   {journal} {Phys. Rev. A}\ }\textbf {\bibinfo {volume} {101}},\ \bibinfo
  {pages} {022333} (\bibinfo {year} {2020})}\BibitemShut {NoStop}%
\bibitem [{\citenamefont {Chen}\ and\ \citenamefont
  {Lucas}(2019)}]{chenFiniteSpeedQuantum2019}%
  \BibitemOpen
  \bibfield  {author} {\bibinfo {author} {\bibfnamefont {C.-F.}\ \bibnamefont
  {Chen}}\ and\ \bibinfo {author} {\bibfnamefont {A.}~\bibnamefont {Lucas}},\
  }\bibfield  {title} {\bibinfo {title} {Finite speed of quantum scrambling
  with long range interactions},\ }\href
  {https://doi.org/10.1103/PhysRevLett.123.250605} {\bibfield  {journal}
  {\bibinfo  {journal} {Phys. Rev. Lett.}\ }\textbf {\bibinfo {volume} {123}},\
  \bibinfo {pages} {250605} (\bibinfo {year} {2019})}\BibitemShut {NoStop}%
\bibitem [{\citenamefont {Tran}\ \emph {et~al.}(2020)\citenamefont {Tran},
  \citenamefont {Chen}, \citenamefont {Ehrenberg}, \citenamefont {Guo},
  \citenamefont {Deshpande}, \citenamefont {Hong}, \citenamefont {Gong},
  \citenamefont {Gorshkov},\ and\ \citenamefont
  {Lucas}}]{tranHierarchyLinearLight2020a}%
  \BibitemOpen
  \bibfield  {author} {\bibinfo {author} {\bibfnamefont {M.~C.}\ \bibnamefont
  {Tran}}, \bibinfo {author} {\bibfnamefont {C.-F.}\ \bibnamefont {Chen}},
  \bibinfo {author} {\bibfnamefont {A.}~\bibnamefont {Ehrenberg}}, \bibinfo
  {author} {\bibfnamefont {A.~Y.}\ \bibnamefont {Guo}}, \bibinfo {author}
  {\bibfnamefont {A.}~\bibnamefont {Deshpande}}, \bibinfo {author}
  {\bibfnamefont {Y.}~\bibnamefont {Hong}}, \bibinfo {author} {\bibfnamefont
  {Z.-X.}\ \bibnamefont {Gong}}, \bibinfo {author} {\bibfnamefont {A.~V.}\
  \bibnamefont {Gorshkov}},\ and\ \bibinfo {author} {\bibfnamefont
  {A.}~\bibnamefont {Lucas}},\ }\bibfield  {title} {\bibinfo {title} {Hierarchy
  of {{Linear Light Cones}} with {{Long}}-{{Range Interactions}}},\ }\href
  {https://doi.org/10.1103/PhysRevX.10.031009} {\bibfield  {journal} {\bibinfo
  {journal} {Phys. Rev. X}\ }\textbf {\bibinfo {volume} {10}},\ \bibinfo
  {pages} {031009} (\bibinfo {year} {2020})}\BibitemShut {NoStop}%
\bibitem [{\citenamefont {Kuwahara}\ and\ \citenamefont
  {Saito}(2020{\natexlab{a}})}]{kuwaharaStrictlyLinearLight2020}%
  \BibitemOpen
  \bibfield  {author} {\bibinfo {author} {\bibfnamefont {T.}~\bibnamefont
  {Kuwahara}}\ and\ \bibinfo {author} {\bibfnamefont {K.}~\bibnamefont
  {Saito}},\ }\bibfield  {title} {\bibinfo {title} {Strictly linear light cones
  in long-range interacting systems of arbitrary dimensions},\ }\href
  {https://doi.org/10.1103/PhysRevX.10.031010} {\bibfield  {journal} {\bibinfo
  {journal} {Phys. Rev. X}\ }\textbf {\bibinfo {volume} {10}},\ \bibinfo
  {pages} {031010} (\bibinfo {year} {2020}{\natexlab{a}})}\BibitemShut
  {NoStop}%
\bibitem [{\citenamefont {Kim}\ \emph {et~al.}(2011{\natexlab{a}})\citenamefont
  {Kim}, \citenamefont {Korenblit}, \citenamefont {Islam}, \citenamefont
  {Edwards}, \citenamefont {Chang}, \citenamefont {Noh}, \citenamefont
  {Carmichael}, \citenamefont {Lin}, \citenamefont {Duan}, \citenamefont
  {Wang}, \citenamefont {Freericks},\ and\ \citenamefont {Monroe}}]{Kim2011}%
  \BibitemOpen
  \bibfield  {author} {\bibinfo {author} {\bibfnamefont {K.}~\bibnamefont
  {Kim}}, \bibinfo {author} {\bibfnamefont {S.}~\bibnamefont {Korenblit}},
  \bibinfo {author} {\bibfnamefont {R.}~\bibnamefont {Islam}}, \bibinfo
  {author} {\bibfnamefont {E.~E.}\ \bibnamefont {Edwards}}, \bibinfo {author}
  {\bibfnamefont {M.-S.}\ \bibnamefont {Chang}}, \bibinfo {author}
  {\bibfnamefont {C.}~\bibnamefont {Noh}}, \bibinfo {author} {\bibfnamefont
  {H.}~\bibnamefont {Carmichael}}, \bibinfo {author} {\bibfnamefont {G.-D.}\
  \bibnamefont {Lin}}, \bibinfo {author} {\bibfnamefont {L.-M.}\ \bibnamefont
  {Duan}}, \bibinfo {author} {\bibfnamefont {C.~C.~J.}\ \bibnamefont {Wang}},
  \bibinfo {author} {\bibfnamefont {J.~K.}\ \bibnamefont {Freericks}},\ and\
  \bibinfo {author} {\bibfnamefont {C.}~\bibnamefont {Monroe}},\ }\bibfield
  {title} {\bibinfo {title} {{Quantum Simulation of the Transverse Ising Model
  With Trapped Ions}},\ }\href
  {http://stacks.iop.org/1367-2630/13/i=10/a=105003} {\bibfield  {journal}
  {\bibinfo  {journal} {New J. Phys.}\ }\textbf {\bibinfo {volume} {13}},\
  \bibinfo {pages} {105003} (\bibinfo {year} {2011}{\natexlab{a}})}\BibitemShut
  {NoStop}%
\bibitem [{\citenamefont {Britton}\ \emph
  {et~al.}(2012{\natexlab{a}})\citenamefont {Britton}, \citenamefont {Sawyer},
  \citenamefont {Keith}, \citenamefont {Wang}, \citenamefont {Freericks},
  \citenamefont {Uys}, \citenamefont {Biercuk},\ and\ \citenamefont
  {Bollinger}}]{britton12}%
  \BibitemOpen
  \bibfield  {author} {\bibinfo {author} {\bibfnamefont {J.~W.}\ \bibnamefont
  {Britton}}, \bibinfo {author} {\bibfnamefont {B.~C.}\ \bibnamefont {Sawyer}},
  \bibinfo {author} {\bibfnamefont {A.~C.}\ \bibnamefont {Keith}}, \bibinfo
  {author} {\bibfnamefont {C.~C.~J.}\ \bibnamefont {Wang}}, \bibinfo {author}
  {\bibfnamefont {J.~K.}\ \bibnamefont {Freericks}}, \bibinfo {author}
  {\bibfnamefont {H.}~\bibnamefont {Uys}}, \bibinfo {author} {\bibfnamefont
  {M.~J.}\ \bibnamefont {Biercuk}},\ and\ \bibinfo {author} {\bibfnamefont
  {J.~J.}\ \bibnamefont {Bollinger}},\ }\bibfield  {title} {\bibinfo {title}
  {Engineered two-dimensional ising interactions in a trapped-ion quantum
  simulator with hundreds of spins},\ }\href
  {https://www.nature.com/articles/nature10981} {\bibfield  {journal} {\bibinfo
   {journal} {Nature}\ }\textbf {\bibinfo {volume} {484}},\ \bibinfo {pages}
  {489} (\bibinfo {year} {2012}{\natexlab{a}})}\BibitemShut {NoStop}%
\bibitem [{Note1()}]{Note1}%
  \BibitemOpen
  \bibinfo {note} {Quantum computing with trapped ions usually uses resonant
  addressing and real excitations of the motional modes~\cite
  {debnath_demonstration_2016}. On the other hand, for one-dimensional chains
  of trapped ions, the off-resonant addressing scheme, which results in spin
  models with tunable approximately-power-law couplings ($\alpha \in [0,3]$),
  is popular among recent analog quantum simulation experiments (For example,
  see Ref.~\cite {Pagano25396}).}\BibitemShut {Stop}%
\bibitem [{\citenamefont {Douglas}\ \emph {et~al.}(2015)\citenamefont
  {Douglas}, \citenamefont {Habibian}, \citenamefont {Hung}, \citenamefont
  {Gorshkov}, \citenamefont {Kimble},\ and\ \citenamefont
  {Chang}}]{Douglas2015}%
  \BibitemOpen
  \bibfield  {author} {\bibinfo {author} {\bibfnamefont {J.~S.}\ \bibnamefont
  {Douglas}}, \bibinfo {author} {\bibfnamefont {H.}~\bibnamefont {Habibian}},
  \bibinfo {author} {\bibfnamefont {C.-L.}\ \bibnamefont {Hung}}, \bibinfo
  {author} {\bibfnamefont {A.~v.}\ \bibnamefont {Gorshkov}}, \bibinfo {author}
  {\bibfnamefont {H.~J.}\ \bibnamefont {Kimble}},\ and\ \bibinfo {author}
  {\bibfnamefont {D.~E.}\ \bibnamefont {Chang}},\ }\bibfield  {title} {\bibinfo
  {title} {{Quantum Many-Body Models With Cold Atoms Coupled to Photonic
  Crystals}},\ }\href {http://dx.doi.org/10.1038/nphoton.2015.57} {\bibfield
  {journal} {\bibinfo  {journal} {Nature Photonics}\ }\textbf {\bibinfo
  {volume} {9}},\ \bibinfo {pages} {326} (\bibinfo {year} {2015})},\ \bibinfo
  {note} {article}\BibitemShut {NoStop}%
\bibitem [{\citenamefont {Gonz{\'a}lez-Tudela}\ \emph
  {et~al.}(2015)\citenamefont {Gonz{\'a}lez-Tudela}, \citenamefont {Hung},
  \citenamefont {Chang}, \citenamefont {Cirac},\ and\ \citenamefont
  {Kimble}}]{gonzalez-tudela15}%
  \BibitemOpen
  \bibfield  {author} {\bibinfo {author} {\bibfnamefont {A.}~\bibnamefont
  {Gonz{\'a}lez-Tudela}}, \bibinfo {author} {\bibfnamefont {C.~L.}\
  \bibnamefont {Hung}}, \bibinfo {author} {\bibfnamefont {D.~E.}\ \bibnamefont
  {Chang}}, \bibinfo {author} {\bibfnamefont {J.~I.}\ \bibnamefont {Cirac}},\
  and\ \bibinfo {author} {\bibfnamefont {H.~J.}\ \bibnamefont {Kimble}},\
  }\bibfield  {title} {\bibinfo {title} {Subwavelength vacuum lattices and
  atom--atom interactions in two-dimensional photonic crystals},\ }\href
  {https://www.nature.com/articles/nphoton.2015.54} {\bibfield  {journal}
  {\bibinfo  {journal} {Nat. Photonics}\ }\textbf {\bibinfo {volume} {9}},\
  \bibinfo {pages} {320} (\bibinfo {year} {2015})}\BibitemShut {NoStop}%
\bibitem [{\citenamefont {Saffman}\ \emph {et~al.}(2010)\citenamefont
  {Saffman}, \citenamefont {Walker},\ and\ \citenamefont
  {M\o{}lmer}}]{Saffman10}%
  \BibitemOpen
  \bibfield  {author} {\bibinfo {author} {\bibfnamefont {M.}~\bibnamefont
  {Saffman}}, \bibinfo {author} {\bibfnamefont {T.~G.}\ \bibnamefont
  {Walker}},\ and\ \bibinfo {author} {\bibfnamefont {K.}~\bibnamefont
  {M\o{}lmer}},\ }\bibfield  {title} {\bibinfo {title} {{Quantum Information
  With Rydberg Atoms}},\ }\href {https://doi.org/10.1103/RevModPhys.82.2313}
  {\bibfield  {journal} {\bibinfo  {journal} {Rev. Mod. Phys.}\ }\textbf
  {\bibinfo {volume} {82}},\ \bibinfo {pages} {2313} (\bibinfo {year}
  {2010})}\BibitemShut {NoStop}%
\bibitem [{\citenamefont {Bernien}\ \emph {et~al.}(2017)\citenamefont
  {Bernien}, \citenamefont {Schwartz}, \citenamefont {Keesling}, \citenamefont
  {Levine}, \citenamefont {Omran}, \citenamefont {Pichler}, \citenamefont
  {Choi}, \citenamefont {Zibrov}, \citenamefont {Endres}, \citenamefont
  {Greiner}, \citenamefont {Vuletic},\ and\ \citenamefont {Lukin}}]{bernien17}%
  \BibitemOpen
  \bibfield  {author} {\bibinfo {author} {\bibfnamefont {H.}~\bibnamefont
  {Bernien}}, \bibinfo {author} {\bibfnamefont {S.}~\bibnamefont {Schwartz}},
  \bibinfo {author} {\bibfnamefont {A.}~\bibnamefont {Keesling}}, \bibinfo
  {author} {\bibfnamefont {H.}~\bibnamefont {Levine}}, \bibinfo {author}
  {\bibfnamefont {A.}~\bibnamefont {Omran}}, \bibinfo {author} {\bibfnamefont
  {H.}~\bibnamefont {Pichler}}, \bibinfo {author} {\bibfnamefont
  {S.}~\bibnamefont {Choi}}, \bibinfo {author} {\bibfnamefont {A.~S.}\
  \bibnamefont {Zibrov}}, \bibinfo {author} {\bibfnamefont {M.}~\bibnamefont
  {Endres}}, \bibinfo {author} {\bibfnamefont {M.}~\bibnamefont {Greiner}},
  \bibinfo {author} {\bibfnamefont {V.}~\bibnamefont {Vuletic}},\ and\ \bibinfo
  {author} {\bibfnamefont {M.~D.}\ \bibnamefont {Lukin}},\ }\bibfield  {title}
  {\bibinfo {title} {Probing many-body dynamics on a 51-atom quantum
  simulator},\ }\href {https://doi.org/10.1038/nature24622} {\bibfield
  {journal} {\bibinfo  {journal} {Nature}\ }\textbf {\bibinfo {volume} {551}},\
  \bibinfo {pages} {579} (\bibinfo {year} {2017})}\BibitemShut {NoStop}%
\bibitem [{\citenamefont {Barredo}\ \emph {et~al.}(2018)\citenamefont
  {Barredo}, \citenamefont {Lienhard}, \citenamefont {de~L{\'e}s{\'e}leuc},
  \citenamefont {Lahaye},\ and\ \citenamefont {Browaeys}}]{barredo18}%
  \BibitemOpen
  \bibfield  {author} {\bibinfo {author} {\bibfnamefont {D.}~\bibnamefont
  {Barredo}}, \bibinfo {author} {\bibfnamefont {V.}~\bibnamefont {Lienhard}},
  \bibinfo {author} {\bibfnamefont {S.}~\bibnamefont {de~L{\'e}s{\'e}leuc}},
  \bibinfo {author} {\bibfnamefont {T.}~\bibnamefont {Lahaye}},\ and\ \bibinfo
  {author} {\bibfnamefont {A.}~\bibnamefont {Browaeys}},\ }\bibfield  {title}
  {\bibinfo {title} {Synthetic three-dimensional atomic structures assembled
  atom by atom},\ }\href {https://www.nature.com/articles/s41586-018-0450-2}
  {\bibfield  {journal} {\bibinfo  {journal} {Nature}\ }\textbf {\bibinfo
  {volume} {561}},\ \bibinfo {pages} {79} (\bibinfo {year} {2018})}\BibitemShut
  {NoStop}%
\bibitem [{\citenamefont {{Maze}}\ \emph {et~al.}(2011)\citenamefont {{Maze}},
  \citenamefont {{Gali}}, \citenamefont {{Togan}}, \citenamefont {{Chu}},
  \citenamefont {{Trifonov}}, \citenamefont {{Kaxiras}},\ and\ \citenamefont
  {{Lukin}}}]{Maze2011}%
  \BibitemOpen
  \bibfield  {author} {\bibinfo {author} {\bibfnamefont {J.~R.}\ \bibnamefont
  {{Maze}}}, \bibinfo {author} {\bibfnamefont {A.}~\bibnamefont {{Gali}}},
  \bibinfo {author} {\bibfnamefont {E.}~\bibnamefont {{Togan}}}, \bibinfo
  {author} {\bibfnamefont {Y.}~\bibnamefont {{Chu}}}, \bibinfo {author}
  {\bibfnamefont {A.}~\bibnamefont {{Trifonov}}}, \bibinfo {author}
  {\bibfnamefont {E.}~\bibnamefont {{Kaxiras}}},\ and\ \bibinfo {author}
  {\bibfnamefont {M.~D.}\ \bibnamefont {{Lukin}}},\ }\bibfield  {title}
  {\bibinfo {title} {{Properties of Nitrogen-Vacancy Centers in Diamond: The
  Group Theoretic Approach}},\ }\href
  {https://doi.org/10.1088/1367-2630/13/2/025025} {\bibfield  {journal}
  {\bibinfo  {journal} {New J. Phys.}\ }\textbf {\bibinfo {volume} {13}},\
  \bibinfo {eid} {025025} (\bibinfo {year} {2011})}\BibitemShut {NoStop}%
\bibitem [{\citenamefont {Yan}\ \emph {et~al.}(2013)\citenamefont {Yan},
  \citenamefont {Moses}, \citenamefont {Gadway}, \citenamefont {Covey},
  \citenamefont {Hazzard}, \citenamefont {Rey}, \citenamefont {Jin},\ and\
  \citenamefont {Ye}}]{yan13}%
  \BibitemOpen
  \bibfield  {author} {\bibinfo {author} {\bibfnamefont {B.}~\bibnamefont
  {Yan}}, \bibinfo {author} {\bibfnamefont {S.~A.}\ \bibnamefont {Moses}},
  \bibinfo {author} {\bibfnamefont {B.}~\bibnamefont {Gadway}}, \bibinfo
  {author} {\bibfnamefont {J.~P.}\ \bibnamefont {Covey}}, \bibinfo {author}
  {\bibfnamefont {K.~R.~A.}\ \bibnamefont {Hazzard}}, \bibinfo {author}
  {\bibfnamefont {A.~M.}\ \bibnamefont {Rey}}, \bibinfo {author} {\bibfnamefont
  {D.~S.}\ \bibnamefont {Jin}},\ and\ \bibinfo {author} {\bibfnamefont
  {J.}~\bibnamefont {Ye}},\ }\bibfield  {title} {\bibinfo {title} {Observation
  of dipolar spin-exchange interactions with lattice-confined polar
  molecules},\ }\href {https://www.nature.com/articles/nature12483} {\bibfield
  {journal} {\bibinfo  {journal} {Nature}\ }\textbf {\bibinfo {volume} {501}},\
  \bibinfo {pages} {521} (\bibinfo {year} {2013})}\BibitemShut {NoStop}%
\bibitem [{\citenamefont {de~L{\'e}s{\'e}leuc}\ \emph
  {et~al.}(2019)\citenamefont {de~L{\'e}s{\'e}leuc}, \citenamefont {Lienhard},
  \citenamefont {Scholl}, \citenamefont {Barredo}, \citenamefont {Weber},
  \citenamefont {Lang}, \citenamefont {B{\"u}chler}, \citenamefont {Lahaye},\
  and\ \citenamefont {Browaeys}}]{leseleuc18b}%
  \BibitemOpen
  \bibfield  {author} {\bibinfo {author} {\bibfnamefont {S.}~\bibnamefont
  {de~L{\'e}s{\'e}leuc}}, \bibinfo {author} {\bibfnamefont {V.}~\bibnamefont
  {Lienhard}}, \bibinfo {author} {\bibfnamefont {P.}~\bibnamefont {Scholl}},
  \bibinfo {author} {\bibfnamefont {D.}~\bibnamefont {Barredo}}, \bibinfo
  {author} {\bibfnamefont {S.}~\bibnamefont {Weber}}, \bibinfo {author}
  {\bibfnamefont {N.}~\bibnamefont {Lang}}, \bibinfo {author} {\bibfnamefont
  {H.~P.}\ \bibnamefont {B{\"u}chler}}, \bibinfo {author} {\bibfnamefont
  {T.}~\bibnamefont {Lahaye}},\ and\ \bibinfo {author} {\bibfnamefont
  {A.}~\bibnamefont {Browaeys}},\ }\bibfield  {title} {\bibinfo {title}
  {Observation of a symmetry-protected topological phase of interacting bosons
  with rydberg atoms},\ }\href {https://doi.org/10.1126/science.aav9105}
  {\bibfield  {journal} {\bibinfo  {journal} {Science}\ }\textbf {\bibinfo
  {volume} {365}},\ \bibinfo {pages} {775} (\bibinfo {year}
  {2019})}\BibitemShut {NoStop}%
\bibitem [{Note2()}]{Note2}%
  \BibitemOpen
  \bibinfo {note} {Because only the total accumulated phase matters in choosing
  the evolution time, we also expect the protocol to be robust against
  experimental errors such as uncertainties in the positions of individual
  particles: If the position of each particle is known up to a precision
  $\varepsilon \ll 1$, the total worst-case error in the accumulated phase
  scales as $t({r_1^{2d}}/{r^\alpha })\times (\varepsilon /r)$, with $r_1$
  being the length of each hypercubes and $r$ being the minimum distance
  between them. The result is a relative phase error proportional to
  $\varepsilon /r$, which becomes smaller and smaller as the distance between
  the hypercubes increases. Moreover, we expect the relative error to be even
  smaller in the commonly occurring situation when uncertainties in the
  positions are uncorrelated between different particles}\BibitemShut {NoStop}%
\bibitem [{\citenamefont {Eldredge}\ \emph {et~al.}(2017)\citenamefont
  {Eldredge}, \citenamefont {Gong}, \citenamefont {Young}, \citenamefont
  {Moosavian}, \citenamefont {Foss-Feig},\ and\ \citenamefont
  {Gorshkov}}]{Eldredge17}%
  \BibitemOpen
  \bibfield  {author} {\bibinfo {author} {\bibfnamefont {Z.}~\bibnamefont
  {Eldredge}}, \bibinfo {author} {\bibfnamefont {Z.-X.}\ \bibnamefont {Gong}},
  \bibinfo {author} {\bibfnamefont {J.~T.}\ \bibnamefont {Young}}, \bibinfo
  {author} {\bibfnamefont {A.~H.}\ \bibnamefont {Moosavian}}, \bibinfo {author}
  {\bibfnamefont {M.}~\bibnamefont {Foss-Feig}},\ and\ \bibinfo {author}
  {\bibfnamefont {A.~V.}\ \bibnamefont {Gorshkov}},\ }\bibfield  {title}
  {\bibinfo {title} {{Fast Quantum State Transfer and Entanglement
  Renormalization Using Long-Range Interactions}},\ }\href
  {https://doi.org/10.1103/PhysRevLett.119.170503} {\bibfield  {journal}
  {\bibinfo  {journal} {Phys. Rev. Lett.}\ }\textbf {\bibinfo {volume} {119}},\
  \bibinfo {pages} {170503} (\bibinfo {year} {2017})}\BibitemShut {NoStop}%
\bibitem [{\citenamefont {Bravyi}\ \emph {et~al.}(2006)\citenamefont {Bravyi},
  \citenamefont {Hastings},\ and\ \citenamefont {Verstraete}}]{Bravyi06}%
  \BibitemOpen
  \bibfield  {author} {\bibinfo {author} {\bibfnamefont {S.}~\bibnamefont
  {Bravyi}}, \bibinfo {author} {\bibfnamefont {M.~B.}\ \bibnamefont
  {Hastings}},\ and\ \bibinfo {author} {\bibfnamefont {F.}~\bibnamefont
  {Verstraete}},\ }\bibfield  {title} {\bibinfo {title} {{Lieb-Robinson Bounds
  and the Generation of Correlations and Topological Quantum Order}},\ }\href
  {https://doi.org/10.1103/PhysRevLett.97.050401} {\bibfield  {journal}
  {\bibinfo  {journal} {Phys. Rev. Lett.}\ }\textbf {\bibinfo {volume} {97}},\
  \bibinfo {pages} {050401} (\bibinfo {year} {2006})}\BibitemShut {NoStop}%
\bibitem [{\citenamefont {Kuwahara}\ and\ \citenamefont
  {Saito}(2020{\natexlab{b}})}]{kuwaharaPolynomialGrowthOutoftimeorder2020a}%
  \BibitemOpen
  \bibfield  {author} {\bibinfo {author} {\bibfnamefont {T.}~\bibnamefont
  {Kuwahara}}\ and\ \bibinfo {author} {\bibfnamefont {K.}~\bibnamefont
  {Saito}},\ }\bibfield  {title} {\bibinfo {title} {Polynomial growth of
  out-of-time-order correlator in arbitrary realistic long-range interacting
  systems},\ }\href {http://arxiv.org/abs/2009.10124} {\bibfield  {journal}
  {\bibinfo  {journal} {arXiv:2009.10124}\ } (\bibinfo {year}
  {2020}{\natexlab{b}})}\BibitemShut {NoStop}%
\bibitem [{\citenamefont {Maze}\ \emph {et~al.}(2008)\citenamefont {Maze},
  \citenamefont {Stanwix}, \citenamefont {Hodges}, \citenamefont {Hong},
  \citenamefont {Taylor}, \citenamefont {Cappellaro}, \citenamefont {Jiang},
  \citenamefont {Dutt}, \citenamefont {Togan}, \citenamefont {Zibrov},
  \citenamefont {Yacoby}, \citenamefont {Walsworth},\ and\ \citenamefont
  {Lukin}}]{mazeNanoscaleMagneticSensing2008}%
  \BibitemOpen
  \bibfield  {author} {\bibinfo {author} {\bibfnamefont {J.~R.}\ \bibnamefont
  {Maze}}, \bibinfo {author} {\bibfnamefont {P.~L.}\ \bibnamefont {Stanwix}},
  \bibinfo {author} {\bibfnamefont {J.~S.}\ \bibnamefont {Hodges}}, \bibinfo
  {author} {\bibfnamefont {S.}~\bibnamefont {Hong}}, \bibinfo {author}
  {\bibfnamefont {J.~M.}\ \bibnamefont {Taylor}}, \bibinfo {author}
  {\bibfnamefont {P.}~\bibnamefont {Cappellaro}}, \bibinfo {author}
  {\bibfnamefont {L.}~\bibnamefont {Jiang}}, \bibinfo {author} {\bibfnamefont
  {M.~V.~G.}\ \bibnamefont {Dutt}}, \bibinfo {author} {\bibfnamefont
  {E.}~\bibnamefont {Togan}}, \bibinfo {author} {\bibfnamefont {A.~S.}\
  \bibnamefont {Zibrov}}, \bibinfo {author} {\bibfnamefont {A.}~\bibnamefont
  {Yacoby}}, \bibinfo {author} {\bibfnamefont {R.~L.}\ \bibnamefont
  {Walsworth}},\ and\ \bibinfo {author} {\bibfnamefont {M.~D.}\ \bibnamefont
  {Lukin}},\ }\bibfield  {title} {{\selectlanguage {en}\bibinfo {title}
  {Nanoscale magnetic sensing with an individual electronic spin in diamond}},\
  }\href {https://doi.org/10.1038/nature07279} {\bibfield  {journal} {\bibinfo
  {journal} {Nature}\ }\textbf {\bibinfo {volume} {455}},\ \bibinfo {pages}
  {644} (\bibinfo {year} {2008})}\BibitemShut {NoStop}%
\bibitem [{\citenamefont {Dolde}\ \emph {et~al.}(2011)\citenamefont {Dolde},
  \citenamefont {Fedder}, \citenamefont {Doherty}, \citenamefont {N{\"o}bauer},
  \citenamefont {Rempp}, \citenamefont {Balasubramanian}, \citenamefont {Wolf},
  \citenamefont {Reinhard}, \citenamefont {Hollenberg}, \citenamefont
  {Jelezko},\ and\ \citenamefont
  {Wrachtrup}}]{doldeSensingElectricFields2011a}%
  \BibitemOpen
  \bibfield  {author} {\bibinfo {author} {\bibfnamefont {F.}~\bibnamefont
  {Dolde}}, \bibinfo {author} {\bibfnamefont {H.}~\bibnamefont {Fedder}},
  \bibinfo {author} {\bibfnamefont {M.~W.}\ \bibnamefont {Doherty}}, \bibinfo
  {author} {\bibfnamefont {T.}~\bibnamefont {N{\"o}bauer}}, \bibinfo {author}
  {\bibfnamefont {F.}~\bibnamefont {Rempp}}, \bibinfo {author} {\bibfnamefont
  {G.}~\bibnamefont {Balasubramanian}}, \bibinfo {author} {\bibfnamefont
  {T.}~\bibnamefont {Wolf}}, \bibinfo {author} {\bibfnamefont {F.}~\bibnamefont
  {Reinhard}}, \bibinfo {author} {\bibfnamefont {L.~C.~L.}\ \bibnamefont
  {Hollenberg}}, \bibinfo {author} {\bibfnamefont {F.}~\bibnamefont
  {Jelezko}},\ and\ \bibinfo {author} {\bibfnamefont {J.}~\bibnamefont
  {Wrachtrup}},\ }\bibfield  {title} {\bibinfo {title} {Sensing electric fields
  using single diamond spins},\ }\href {https://doi.org/10.1038/nphys1969}
  {\bibfield  {journal} {\bibinfo  {journal} {Nature Phys.}\ }\textbf {\bibinfo
  {volume} {7}},\ \bibinfo {pages} {459} (\bibinfo {year} {2011})}\BibitemShut
  {NoStop}%
\bibitem [{\citenamefont {Sedlacek}\ \emph {et~al.}(2012)\citenamefont
  {Sedlacek}, \citenamefont {Schwettmann}, \citenamefont {K{\"u}bler},
  \citenamefont {L{\"o}w}, \citenamefont {Pfau},\ and\ \citenamefont
  {Shaffer}}]{sedlacekMicrowaveElectrometryRydberg2012}%
  \BibitemOpen
  \bibfield  {author} {\bibinfo {author} {\bibfnamefont {J.~A.}\ \bibnamefont
  {Sedlacek}}, \bibinfo {author} {\bibfnamefont {A.}~\bibnamefont
  {Schwettmann}}, \bibinfo {author} {\bibfnamefont {H.}~\bibnamefont
  {K{\"u}bler}}, \bibinfo {author} {\bibfnamefont {R.}~\bibnamefont {L{\"o}w}},
  \bibinfo {author} {\bibfnamefont {T.}~\bibnamefont {Pfau}},\ and\ \bibinfo
  {author} {\bibfnamefont {J.~P.}\ \bibnamefont {Shaffer}},\ }\bibfield
  {title} {{\selectlanguage {en}\bibinfo {title} {Microwave electrometry with
  {{Rydberg}} atoms in a vapour cell using bright atomic resonances}},\ }\href
  {https://doi.org/10.1038/nphys2423} {\bibfield  {journal} {\bibinfo
  {journal} {Nature Physics}\ }\textbf {\bibinfo {volume} {8}},\ \bibinfo
  {pages} {819} (\bibinfo {year} {2012})}\BibitemShut {NoStop}%
\bibitem [{\citenamefont {Wade}\ \emph {et~al.}(2017)\citenamefont {Wade},
  \citenamefont {{\v S}ibali{\'c}}, \citenamefont {{de Melo}}, \citenamefont
  {Kondo}, \citenamefont {Adams},\ and\ \citenamefont
  {Weatherill}}]{wadeRealTimeNearFieldTerahertz2017}%
  \BibitemOpen
  \bibfield  {author} {\bibinfo {author} {\bibfnamefont {C.~G.}\ \bibnamefont
  {Wade}}, \bibinfo {author} {\bibfnamefont {N.}~\bibnamefont {{\v
  S}ibali{\'c}}}, \bibinfo {author} {\bibfnamefont {N.~R.}\ \bibnamefont {{de
  Melo}}}, \bibinfo {author} {\bibfnamefont {J.~M.}\ \bibnamefont {Kondo}},
  \bibinfo {author} {\bibfnamefont {C.~S.}\ \bibnamefont {Adams}},\ and\
  \bibinfo {author} {\bibfnamefont {K.~J.}\ \bibnamefont {Weatherill}},\
  }\bibfield  {title} {\bibinfo {title} {Real-{{Time Near}}-{{Field Terahertz
  Imaging}} with {{Atomic Optical Fluorescence}}},\ }\href
  {https://doi.org/10.1038/nphoton.2016.214} {\bibfield  {journal} {\bibinfo
  {journal} {Nat. Photonics}\ }\textbf {\bibinfo {volume} {11}},\ \bibinfo
  {pages} {40} (\bibinfo {year} {2017})}\BibitemShut {NoStop}%
\bibitem [{\citenamefont {Carr}\ \emph {et~al.}(2009)\citenamefont {Carr},
  \citenamefont {DeMille}, \citenamefont {Krems},\ and\ \citenamefont
  {Ye}}]{carrColdUltracoldMolecules2009}%
  \BibitemOpen
  \bibfield  {author} {\bibinfo {author} {\bibfnamefont {L.~D.}\ \bibnamefont
  {Carr}}, \bibinfo {author} {\bibfnamefont {D.}~\bibnamefont {DeMille}},
  \bibinfo {author} {\bibfnamefont {R.~V.}\ \bibnamefont {Krems}},\ and\
  \bibinfo {author} {\bibfnamefont {J.}~\bibnamefont {Ye}},\ }\bibfield
  {title} {{\selectlanguage {en}\bibinfo {title} {Cold and ultracold molecules:
  Science, technology and applications}},\ }\href
  {https://doi.org/10.1088/1367-2630/11/5/055049} {\bibfield  {journal}
  {\bibinfo  {journal} {New J. Phys.}\ }\textbf {\bibinfo {volume} {11}},\
  \bibinfo {pages} {055049} (\bibinfo {year} {2009})}\BibitemShut {NoStop}%
\bibitem [{\citenamefont {Andr{\'e}}\ \emph {et~al.}(2004)\citenamefont
  {Andr{\'e}}, \citenamefont {S{\o}rensen},\ and\ \citenamefont
  {Lukin}}]{andreStabilityAtomicClocks2004}%
  \BibitemOpen
  \bibfield  {author} {\bibinfo {author} {\bibfnamefont {A.}~\bibnamefont
  {Andr{\'e}}}, \bibinfo {author} {\bibfnamefont {A.~S.}\ \bibnamefont
  {S{\o}rensen}},\ and\ \bibinfo {author} {\bibfnamefont {M.~D.}\ \bibnamefont
  {Lukin}},\ }\bibfield  {title} {\bibinfo {title} {Stability of {{Atomic
  Clocks Based}} on {{Entangled Atoms}}},\ }\href
  {https://doi.org/10.1103/PhysRevLett.92.230801} {\bibfield  {journal}
  {\bibinfo  {journal} {Phys. Rev. Lett.}\ }\textbf {\bibinfo {volume} {92}},\
  \bibinfo {pages} {230801} (\bibinfo {year} {2004})}\BibitemShut {NoStop}%
\bibitem [{\citenamefont {Britton}\ \emph
  {et~al.}(2012{\natexlab{b}})\citenamefont {Britton}, \citenamefont {Sawyer},
  \citenamefont {Keith}, \citenamefont {Wang}, \citenamefont {Freericks},
  \citenamefont {Uys}, \citenamefont {Biercuk},\ and\ \citenamefont
  {Bollinger}}]{brittonEngineeredTwodimensionalIsing2012}%
  \BibitemOpen
  \bibfield  {author} {\bibinfo {author} {\bibfnamefont {J.~W.}\ \bibnamefont
  {Britton}}, \bibinfo {author} {\bibfnamefont {B.~C.}\ \bibnamefont {Sawyer}},
  \bibinfo {author} {\bibfnamefont {A.~C.}\ \bibnamefont {Keith}}, \bibinfo
  {author} {\bibfnamefont {C.-C.~J.}\ \bibnamefont {Wang}}, \bibinfo {author}
  {\bibfnamefont {J.~K.}\ \bibnamefont {Freericks}}, \bibinfo {author}
  {\bibfnamefont {H.}~\bibnamefont {Uys}}, \bibinfo {author} {\bibfnamefont
  {M.~J.}\ \bibnamefont {Biercuk}},\ and\ \bibinfo {author} {\bibfnamefont
  {J.~J.}\ \bibnamefont {Bollinger}},\ }\bibfield  {title} {{\selectlanguage
  {en}\bibinfo {title} {Engineered two-dimensional {{Ising}} interactions in a
  trapped-ion quantum simulator with hundreds of spins}},\ }\href
  {https://doi.org/10.1038/nature10981} {\bibfield  {journal} {\bibinfo
  {journal} {Nature}\ }\textbf {\bibinfo {volume} {484}},\ \bibinfo {pages}
  {489} (\bibinfo {year} {2012}{\natexlab{b}})}\BibitemShut {NoStop}%
\bibitem [{\citenamefont {Kim}\ \emph {et~al.}(2011{\natexlab{b}})\citenamefont
  {Kim}, \citenamefont {Korenblit}, \citenamefont {Islam}, \citenamefont
  {Edwards}, \citenamefont {Chang}, \citenamefont {Noh}, \citenamefont
  {Carmichael}, \citenamefont {Lin}, \citenamefont {Duan}, \citenamefont
  {Wang}, \citenamefont {Freericks},\ and\ \citenamefont
  {Monroe}}]{kimQuantumSimulationTransverse2011}%
  \BibitemOpen
  \bibfield  {author} {\bibinfo {author} {\bibfnamefont {K.}~\bibnamefont
  {Kim}}, \bibinfo {author} {\bibfnamefont {S.}~\bibnamefont {Korenblit}},
  \bibinfo {author} {\bibfnamefont {R.}~\bibnamefont {Islam}}, \bibinfo
  {author} {\bibfnamefont {E.~E.}\ \bibnamefont {Edwards}}, \bibinfo {author}
  {\bibfnamefont {M.-S.}\ \bibnamefont {Chang}}, \bibinfo {author}
  {\bibfnamefont {C.}~\bibnamefont {Noh}}, \bibinfo {author} {\bibfnamefont
  {H.}~\bibnamefont {Carmichael}}, \bibinfo {author} {\bibfnamefont {G.-D.}\
  \bibnamefont {Lin}}, \bibinfo {author} {\bibfnamefont {L.-M.}\ \bibnamefont
  {Duan}}, \bibinfo {author} {\bibfnamefont {C.~C.~J.}\ \bibnamefont {Wang}},
  \bibinfo {author} {\bibfnamefont {J.~K.}\ \bibnamefont {Freericks}},\ and\
  \bibinfo {author} {\bibfnamefont {C.}~\bibnamefont {Monroe}},\ }\bibfield
  {title} {{\selectlanguage {en}\bibinfo {title} {Quantum simulation of the
  transverse {{Ising}} model with trapped ions}},\ }\href
  {https://doi.org/10.1088/1367-2630/13/10/105003} {\bibfield  {journal}
  {\bibinfo  {journal} {New J. Phys.}\ }\textbf {\bibinfo {volume} {13}},\
  \bibinfo {pages} {105003} (\bibinfo {year} {2011}{\natexlab{b}})}\BibitemShut
  {NoStop}%
\bibitem [{\citenamefont {H{\o}yer}\ and\ \citenamefont {{\v
  S}palek}(2005)}]{hoyerQuantumFanoutPowerful2005}%
  \BibitemOpen
  \bibfield  {author} {\bibinfo {author} {\bibfnamefont {P.}~\bibnamefont
  {H{\o}yer}}\ and\ \bibinfo {author} {\bibfnamefont {R.}~\bibnamefont {{\v
  S}palek}},\ }\bibfield  {title} {{\selectlanguage {EN}\bibinfo {title}
  {Quantum {{Fan}}-out is {{Powerful}}}},\ }\href
  {https://doi.org/10.4086/toc.2005.v001a005} {\bibfield  {journal} {\bibinfo
  {journal} {Theory Comput.}\ }\textbf {\bibinfo {volume} {1}},\ \bibinfo
  {pages} {81} (\bibinfo {year} {2005})}\BibitemShut {NoStop}%
\bibitem [{\citenamefont {Aguado}\ and\ \citenamefont
  {Vidal}(2008)}]{aguadoEntanglementRenormalizationTopological2008}%
  \BibitemOpen
  \bibfield  {author} {\bibinfo {author} {\bibfnamefont {M.}~\bibnamefont
  {Aguado}}\ and\ \bibinfo {author} {\bibfnamefont {G.}~\bibnamefont {Vidal}},\
  }\bibfield  {title} {\bibinfo {title} {Entanglement {{Renormalization}} and
  {{Topological Order}}},\ }\href
  {https://doi.org/10.1103/PhysRevLett.100.070404} {\bibfield  {journal}
  {\bibinfo  {journal} {Phys. Rev. Lett.}\ }\textbf {\bibinfo {volume} {100}},\
  \bibinfo {pages} {070404} (\bibinfo {year} {2008})}\BibitemShut {NoStop}%
\bibitem [{\citenamefont {Vidal}(2007)}]{vidalEntanglementRenormalization2007}%
  \BibitemOpen
  \bibfield  {author} {\bibinfo {author} {\bibfnamefont {G.}~\bibnamefont
  {Vidal}},\ }\bibfield  {title} {\bibinfo {title} {Entanglement
  {{Renormalization}}},\ }\href {https://doi.org/10.1103/PhysRevLett.99.220405}
  {\bibfield  {journal} {\bibinfo  {journal} {Phys. Rev. Lett.}\ }\textbf
  {\bibinfo {volume} {99}},\ \bibinfo {pages} {220405} (\bibinfo {year}
  {2007})}\BibitemShut {NoStop}%
\bibitem [{\citenamefont {Vidal}(2008)}]{vidalClassQuantumManyBody2008}%
  \BibitemOpen
  \bibfield  {author} {\bibinfo {author} {\bibfnamefont {G.}~\bibnamefont
  {Vidal}},\ }\bibfield  {title} {\bibinfo {title} {Class of {{Quantum
  Many}}-{{Body States That Can Be Efficiently Simulated}}},\ }\href
  {https://doi.org/10.1103/PhysRevLett.101.110501} {\bibfield  {journal}
  {\bibinfo  {journal} {Phys. Rev. Lett.}\ }\textbf {\bibinfo {volume} {101}},\
  \bibinfo {pages} {110501} (\bibinfo {year} {2008})}\BibitemShut {NoStop}%
\bibitem [{\citenamefont {Giovannetti}\ \emph {et~al.}(2008)\citenamefont
  {Giovannetti}, \citenamefont {Montangero},\ and\ \citenamefont
  {Fazio}}]{giovannettiQuantumMultiscaleEntanglement2008}%
  \BibitemOpen
  \bibfield  {author} {\bibinfo {author} {\bibfnamefont {V.}~\bibnamefont
  {Giovannetti}}, \bibinfo {author} {\bibfnamefont {S.}~\bibnamefont
  {Montangero}},\ and\ \bibinfo {author} {\bibfnamefont {R.}~\bibnamefont
  {Fazio}},\ }\bibfield  {title} {\bibinfo {title} {Quantum {{Multiscale
  Entanglement Renormalization Ansatz Channels}}},\ }\href
  {https://doi.org/10.1103/PhysRevLett.101.180503} {\bibfield  {journal}
  {\bibinfo  {journal} {Phys. Rev. Lett.}\ }\textbf {\bibinfo {volume} {101}},\
  \bibinfo {pages} {180503} (\bibinfo {year} {2008})}\BibitemShut {NoStop}%
\bibitem [{\citenamefont {Tran}\ \emph
  {et~al.}(2019{\natexlab{b}})\citenamefont {Tran}, \citenamefont {Ehrenberg},
  \citenamefont {Guo}, \citenamefont {Titum}, \citenamefont {Abanin},\ and\
  \citenamefont {Gorshkov}}]{tranLocalityHeatingPeriodically2019}%
  \BibitemOpen
  \bibfield  {author} {\bibinfo {author} {\bibfnamefont {M.~C.}\ \bibnamefont
  {Tran}}, \bibinfo {author} {\bibfnamefont {A.}~\bibnamefont {Ehrenberg}},
  \bibinfo {author} {\bibfnamefont {A.~Y.}\ \bibnamefont {Guo}}, \bibinfo
  {author} {\bibfnamefont {P.}~\bibnamefont {Titum}}, \bibinfo {author}
  {\bibfnamefont {D.~A.}\ \bibnamefont {Abanin}},\ and\ \bibinfo {author}
  {\bibfnamefont {A.~V.}\ \bibnamefont {Gorshkov}},\ }\bibfield  {title}
  {\bibinfo {title} {Locality and heating in periodically driven,
  power-law-interacting systems},\ }\href
  {https://doi.org/10.1103/PhysRevA.100.052103} {\bibfield  {journal} {\bibinfo
   {journal} {Phys. Rev. A}\ }\textbf {\bibinfo {volume} {100}},\ \bibinfo
  {pages} {052103} (\bibinfo {year} {2019}{\natexlab{b}})}\BibitemShut
  {NoStop}%
\bibitem [{\citenamefont {Eisert}\ \emph {et~al.}(2013)\citenamefont {Eisert},
  \citenamefont {{van den Worm}}, \citenamefont {Manmana},\ and\ \citenamefont
  {Kastner}}]{eisertBreakdownQuasilocalityLongRange2013}%
  \BibitemOpen
  \bibfield  {author} {\bibinfo {author} {\bibfnamefont {J.}~\bibnamefont
  {Eisert}}, \bibinfo {author} {\bibfnamefont {M.}~\bibnamefont {{van den
  Worm}}}, \bibinfo {author} {\bibfnamefont {S.~R.}\ \bibnamefont {Manmana}},\
  and\ \bibinfo {author} {\bibfnamefont {M.}~\bibnamefont {Kastner}},\
  }\bibfield  {title} {\bibinfo {title} {Breakdown of {{Quasilocality}} in
  {{Long}}-{{Range Quantum Lattice Models}}},\ }\href
  {https://doi.org/10.1103/PhysRevLett.111.260401} {\bibfield  {journal}
  {\bibinfo  {journal} {Phys. Rev. Lett.}\ }\textbf {\bibinfo {volume} {111}},\
  \bibinfo {pages} {260401} (\bibinfo {year} {2013})}\BibitemShut {NoStop}%
\bibitem [{\citenamefont {Hauke}\ and\ \citenamefont
  {Tagliacozzo}(2013)}]{haukeSpreadCorrelationsLongRange2013a}%
  \BibitemOpen
  \bibfield  {author} {\bibinfo {author} {\bibfnamefont {P.}~\bibnamefont
  {Hauke}}\ and\ \bibinfo {author} {\bibfnamefont {L.}~\bibnamefont
  {Tagliacozzo}},\ }\bibfield  {title} {\bibinfo {title} {Spread of
  {{Correlations}} in {{Long}}-{{Range Interacting Quantum Systems}}},\ }\href
  {https://doi.org/10.1103/PhysRevLett.111.207202} {\bibfield  {journal}
  {\bibinfo  {journal} {Phys. Rev. Lett.}\ }\textbf {\bibinfo {volume} {111}},\
  \bibinfo {pages} {207202} (\bibinfo {year} {2013})}\BibitemShut {NoStop}%
\bibitem [{\citenamefont {Guo}\ \emph {et~al.}(2020{\natexlab{b}})\citenamefont
  {Guo}, \citenamefont {Tran}, \citenamefont {Childs}, \citenamefont
  {Gorshkov},\ and\ \citenamefont
  {Gong}}]{guoSignalingScramblingStrongly2020a}%
  \BibitemOpen
  \bibfield  {author} {\bibinfo {author} {\bibfnamefont {A.~Y.}\ \bibnamefont
  {Guo}}, \bibinfo {author} {\bibfnamefont {M.~C.}\ \bibnamefont {Tran}},
  \bibinfo {author} {\bibfnamefont {A.~M.}\ \bibnamefont {Childs}}, \bibinfo
  {author} {\bibfnamefont {A.~V.}\ \bibnamefont {Gorshkov}},\ and\ \bibinfo
  {author} {\bibfnamefont {Z.-X.}\ \bibnamefont {Gong}},\ }\bibfield  {title}
  {\bibinfo {title} {Signaling and {{Scrambling}} with {{Strongly
  Long}}-{{Range Interactions}}},\ }\href
  {https://doi.org/10.1103/PhysRevA.102.010401} {\bibfield  {journal} {\bibinfo
   {journal} {Phys. Rev. A}\ }\textbf {\bibinfo {volume} {102}},\ \bibinfo
  {pages} {010401} (\bibinfo {year} {2020}{\natexlab{b}})}\BibitemShut
  {NoStop}%
\bibitem [{\citenamefont {Haah}\ \emph {et~al.}(2018)\citenamefont {Haah},
  \citenamefont {Hastings}, \citenamefont {Kothari},\ and\ \citenamefont
  {Low}}]{Haah}%
  \BibitemOpen
  \bibfield  {author} {\bibinfo {author} {\bibfnamefont {J.}~\bibnamefont
  {Haah}}, \bibinfo {author} {\bibfnamefont {M.~B.}\ \bibnamefont {Hastings}},
  \bibinfo {author} {\bibfnamefont {R.}~\bibnamefont {Kothari}},\ and\ \bibinfo
  {author} {\bibfnamefont {G.~H.}\ \bibnamefont {Low}},\ }\bibfield  {title}
  {\bibinfo {title} {Quantum algorithm for simulating real time evolution of
  lattice hamiltonians},\ }in\ \href {https://doi.org/10.1109/FOCS.2018.00041}
  {\emph {\bibinfo {booktitle} {59th {IEEE} Annual Symposium on Foundations of
  Computer Science, {FOCS} 2018, Paris, France, October 7-9, 2018}}},\ \bibinfo
  {editor} {edited by\ \bibinfo {editor} {\bibfnamefont {M.}~\bibnamefont
  {Thorup}}}\ (\bibinfo  {publisher} {{IEEE} Computer Society},\ \bibinfo
  {year} {2018})\ pp.\ \bibinfo {pages} {350--360}\BibitemShut {NoStop}%
\bibitem [{\citenamefont {Childs}\ \emph {et~al.}(2019)\citenamefont {Childs},
  \citenamefont {Su}, \citenamefont {Tran}, \citenamefont {Wiebe},\ and\
  \citenamefont {Zhu}}]{Childs2019d}%
  \BibitemOpen
  \bibfield  {author} {\bibinfo {author} {\bibfnamefont {A.~M.}\ \bibnamefont
  {Childs}}, \bibinfo {author} {\bibfnamefont {Y.}~\bibnamefont {Su}}, \bibinfo
  {author} {\bibfnamefont {M.~C.}\ \bibnamefont {Tran}}, \bibinfo {author}
  {\bibfnamefont {N.}~\bibnamefont {Wiebe}},\ and\ \bibinfo {author}
  {\bibfnamefont {S.}~\bibnamefont {Zhu}},\ }\bibfield  {title} {\bibinfo
  {title} {A theory of trotter error},\ }\href {arXiv:1912.08854} {\bibfield
  {journal} {\bibinfo  {journal} {arXiv:1912.08854}\ } (\bibinfo {year}
  {2019})}\BibitemShut {NoStop}%
\bibitem [{\citenamefont {Debnath}\ \emph {et~al.}(2016)\citenamefont
  {Debnath}, \citenamefont {Linke}, \citenamefont {Figgatt}, \citenamefont
  {Landsman}, \citenamefont {Wright},\ and\ \citenamefont
  {Monroe}}]{debnath_demonstration_2016}%
  \BibitemOpen
  \bibfield  {author} {\bibinfo {author} {\bibfnamefont {S.}~\bibnamefont
  {Debnath}}, \bibinfo {author} {\bibfnamefont {N.~M.}\ \bibnamefont {Linke}},
  \bibinfo {author} {\bibfnamefont {C.}~\bibnamefont {Figgatt}}, \bibinfo
  {author} {\bibfnamefont {K.~A.}\ \bibnamefont {Landsman}}, \bibinfo {author}
  {\bibfnamefont {K.}~\bibnamefont {Wright}},\ and\ \bibinfo {author}
  {\bibfnamefont {C.}~\bibnamefont {Monroe}},\ }\bibfield  {title}
  {{\selectlanguage {en}\bibinfo {title} {Demonstration of a small programmable
  quantum computer with atomic qubits}},\ }\href
  {https://doi.org/10.1038/nature18648} {\bibfield  {journal} {\bibinfo
  {journal} {Nature}\ }\textbf {\bibinfo {volume} {536}},\ \bibinfo {pages}
  {63} (\bibinfo {year} {2016})}\BibitemShut {NoStop}%
\bibitem [{\citenamefont {Pagano}\ \emph {et~al.}(2020)\citenamefont {Pagano},
  \citenamefont {Bapat}, \citenamefont {Becker}, \citenamefont {Collins},
  \citenamefont {De}, \citenamefont {Hess}, \citenamefont {Kaplan},
  \citenamefont {Kyprianidis}, \citenamefont {Tan}, \citenamefont {Baldwin},
  \citenamefont {Brady}, \citenamefont {Deshpande}, \citenamefont {Liu},
  \citenamefont {Jordan}, \citenamefont {Gorshkov},\ and\ \citenamefont
  {Monroe}}]{Pagano25396}%
  \BibitemOpen
  \bibfield  {author} {\bibinfo {author} {\bibfnamefont {G.}~\bibnamefont
  {Pagano}}, \bibinfo {author} {\bibfnamefont {A.}~\bibnamefont {Bapat}},
  \bibinfo {author} {\bibfnamefont {P.}~\bibnamefont {Becker}}, \bibinfo
  {author} {\bibfnamefont {K.~S.}\ \bibnamefont {Collins}}, \bibinfo {author}
  {\bibfnamefont {A.}~\bibnamefont {De}}, \bibinfo {author} {\bibfnamefont
  {P.~W.}\ \bibnamefont {Hess}}, \bibinfo {author} {\bibfnamefont {H.~B.}\
  \bibnamefont {Kaplan}}, \bibinfo {author} {\bibfnamefont {A.}~\bibnamefont
  {Kyprianidis}}, \bibinfo {author} {\bibfnamefont {W.~L.}\ \bibnamefont
  {Tan}}, \bibinfo {author} {\bibfnamefont {C.}~\bibnamefont {Baldwin}},
  \bibinfo {author} {\bibfnamefont {L.~T.}\ \bibnamefont {Brady}}, \bibinfo
  {author} {\bibfnamefont {A.}~\bibnamefont {Deshpande}}, \bibinfo {author}
  {\bibfnamefont {F.}~\bibnamefont {Liu}}, \bibinfo {author} {\bibfnamefont
  {S.}~\bibnamefont {Jordan}}, \bibinfo {author} {\bibfnamefont {A.~V.}\
  \bibnamefont {Gorshkov}},\ and\ \bibinfo {author} {\bibfnamefont
  {C.}~\bibnamefont {Monroe}},\ }\bibfield  {title} {\bibinfo {title} {Quantum
  approximate optimization of the long-range ising model with a trapped-ion
  quantum simulator},\ }\href {https://doi.org/10.1073/pnas.2006373117}
  {\bibfield  {journal} {\bibinfo  {journal} {Proc. Natl. Acad. Sci.}\ }\textbf
  {\bibinfo {volume} {117}},\ \bibinfo {pages} {25396} (\bibinfo {year}
  {2020})}\BibitemShut {NoStop}%
\end{thebibliography}%
%%%%%%%%%%%%%%%%%%%%%%%%%%%%%%%%%%%%%%%%%%%%%
\end{document}